\documentclass[a4paper,11pt]{article}

\usepackage{amssymb}
\usepackage{wasysym}
\usepackage{dsfont}
\usepackage{xytree}
\usepackage{stmaryrd}
\usepackage{rotating}

\title{A formula for the fractal dimension $d \sim 0.87$ of the Cantorian set underlying the Devil's staircase associated with the Circle Map}
\author{M. N. Piacquadio Losada \\
Facultad de Ingenier\'{i}a. Universidad de Buenos Aires\\
Paseo Col\'{o}n 850, (1063) Buenos Aires, Argentina}
\begin{document}
\maketitle
\begin{abstract}
The Cantor set complementary to the Devil's Staircase associated with the Circle Map has a fractal dimension $d \sim 0.87$, universal for a wide range of maps, such results being of a numerical character. In this paper we deduce a formula for such dimensional value, the corresponding theoretical reasoning permits conjecturing on the nature of its universality. The Devil's Staircase associated with the Circle Map is a function that transforms horizontal unit interval $I$ onto $I$, and is endowed with the Farey-Brocot $(F-B)$ structure in the vertical axis via the rational heights of stability intervals. The underlying Cantordust fractal set $\Omega$ in the horizontal axis, $\Omega \subset I$, with fractal dimension $d(\Omega) = d \sim 0.87$ has a natural covering with segments that also follow the $F-B$ hierarchy: the staircase associates vertical $I$ (of unit dimension) with horizontal $\Omega \subset I$ (of dimension $\sim 0.87$), i.e. it selects a certain subset $\Omega$ of $I$, both sets $F-B$ structured, $\Omega$ with smaller dimension than that of $I$. Hence, the structure of the staircase mirrors the $F-B$ hierarchy. In this paper we consider the subset $\Omega_{F-B}$ of $I$ that concentrates the measure induced by the $F-B$ partition and calculate its Hausdorff dimension, i.e. the entropic or information dimension of the $F-B$ measure, and show that it coincides with $d(\Omega) \sim 0.87$. Hence, this dimensional value stems from the $F-B$ structure, and we draw conclusions and conjectures from this fact. Finally, we calculate the statistical "Euclidean" dimension (based on the ordinary Lebesgue measure) of the $F-B$ partition, and we show that it is the same as $d(\Omega_{F-B})$, which permits conjecturing on the universality of the dimensional value $d \sim 0.87$.
\end{abstract}
\section{Introduction}
The Cantor set complementary to the Devil's Staircase associated with the Circle Map has a fractal dimension $d \sim 0.87$ [Jensen et al., 1984], universal for a wide range of maps [Bak, 1986], such results being of a numerical character. In this paper we deduce a formula for such dimensional value, the corresponding theoretical reasoning permits conjecturing on the nature of its universality.
\newline\indent
A Cantor or Devil's staircase is an increasing function $y = g(x)$ from $I = [0,1]$ onto $I$, with zero derivative almost everywhere, constant in the so-called intervals of resonance or stability $\Delta x_k$, $k \in \mathds{N}$, which are infinite in number. Such staircases are frequently observed in empirical physics [Bak, 1986], and their universal properties are of great interest. The complement in $I$ of $U_{k \in \mathds{N}}\Delta x_k$ is a totally discontinuous Cantor-dust set $\Omega$ naturally associated with the staircase, which reflects the features of the particular physical problem under study. The sine circle map $\theta_{n+1}=\theta_n + w + \frac{1}{2\pi} \sin ( 2\pi\theta_n )$ is a simple model describing [Bak, 1986] systems with two competing frequencies, e.g. the forced pendulum, with $\theta$ the angle formed by the vertical and the pendulum, $n$ the discretized time variable, and $w$ the frequency of the system in the absence of the non-linear term given by the sine function. Let $W = \lim_{n \rightarrow \infty} \frac{\theta_n}{n}$ be the winding number of the system. The graph of the function $W = g(w)$ is a well known Cantor staircase; with $\Delta w$ we denote an interval of stability as well as the corresponding stair step. 
\newline\indent
Let $\Delta w_1$ and $\Delta w_2$ be two such resonance intervals such that all intervals between these two have smaller length. Let $g(x) = \frac{a}{b}$ if $x \in \Delta w_1$, $g(x) = \frac{a'}{b'}$ if $x \in \Delta w_2$, all stair steps have rational height. If $x$ is in the largest interval in the gap between $\Delta w_1$ and $\Delta w_2$, then $g(x) = \frac{a+a'}{b+b'}$; i.e. the height of stair steps follows what, by definition, is the Farey-Brocot $(F-B)$ interpolation law. This is so for many staircases empirically found in physics and other sciences. Starting from $\frac{a}{b} = \frac{0}{1}= 0$, and $\frac{a'}{b'} =\frac{1}{1} = 1$,  $F-B$ interpolates $\frac{0+1}{1+1} = \frac{1}{2}$ between $0$ and $1$, partitioning $I$ in two intervals, in turn partitioned in two intervals each, yielding a partition of $I$ in $4$ intervals in the second order interpolation\dots and in $2^N$ intervals in the $N^{th}$ order of the $F-B$ interpolation. The induced $F-B$ measure in level $N$ of interpolation gives the same probability measure \textemdash by definition\textemdash\space i.e. $\frac{1}{2^N}$, to each of these $2^N$ intervals. Let $C_N = I - U_{\frac{p}{q}}\Delta w_{\frac{p}{q}}$ be such that $w_{\frac{p}{q}}$ is the stair step of height $\frac{p}{q}$, and $\frac{p}{q}$ is every rational in the $N^{th}$ level of $F-B$ interpolation. Then $C_N$ is a covering of $\Omega$ by $2^N$ intervals $I^{(N)}_j$ in the horizontal $w$ axis, such that $g(I_j^{(N)})$, $j:1 \rightarrow 2^N$ are the $2^N$ intervals of the $F-B$ partition of $I$ in the vertical $W$ axis. If we plot [Piacquadio, 2004] lengths of the $2^N$ intervals $I_j^{(N)}$ against length of the corresponding $g(I_j^{(N)})$ we obtain a straight line that passes through the origin with slope $c_N$ growing as $N$ grows. So $\Omega$ also follows the hierarchy of the Farey tree via its covering: the staircase relates an $F-B$ structured unit segment $I$ (dimension $1$) with an $F-B$ structured subset $\Omega$ (dimension $ \sim 0.870$ ) of $I$, i.e. the staircase selects a subset of $I$ of smaller dimension, the $F-B$ partition being at the core of the very structure of the staircase. Hence, it seems natural to relate the $F-B$ partition to the $\sim 0.870$ value: using the tools of multifractality, we calculate the multifractal spectrum $(\alpha,f(\alpha))$ of the $F-B$ measure on $I$, and identify which subfractal $\Omega_{F-B} \subset I$ has a dimension $\sim 0.870$. We find that $\Omega_{F-B}$ is the set that concentrates the $F-B$ measure, it corresponds to the value $\alpha$ for which $f(\alpha) = \alpha$ and $f'(\alpha) = 1$, i.e. its dimension $\sim 0.870$ is the entropic or information dimension of the $F-B$ measure.
\newline\indent
A feature in the importance of this dimensional value is its universality, which has been checked (again [Bak, 1986])  by studying a broad class of circle maps with more complicated non-linear terms than the simple sine map. Although details may differ from those of the sine map $W=g(w)$ \textemdash steps narrower, sometimes larger\textemdash\space still the dimension of the underlying  $\Omega$ remains $\sim 0.870$.
\newline\indent
We proceed as follows: the $2^N$ intervals in the $N^{th}$ $F-B$ partition have the same $F-B$ measure, $\frac{1}{2^N}$, but very different lengths, i.e. very different Euclidean measure. We start (Secs. 3 and 4) with the thermodynamical algorithm $(\alpha,f(\alpha))$ for the hereinafter called Euclidean case, and by this we mean: all segments considered have equal Euclidean length at any  $N^{th}$ partition. We proceed from there in slow steps in such a way that the results can be extended to the $F-B$ measure (Sec. 5) in a manner that \textemdash we trust\textemdash\space will be seen as "natural". Thus, a first connection between the two measures will be established: working always in $I$, we express $\alpha$ and $f(\alpha)$ in terms of contractors (probability contractions $p_j \in (0,1)$ and/or length contractions $c_j \in (0,1)$) and their key frequencies linked to each other through the thermodynamical algorithm; a finite number of contractors for the so-called Euclidean case, extending the results to an infinite number of contractors in the $F-B$ case. Next, we estimate (Sec. 5.5) the Hausdorff dimension of the subfractal $\Omega_{F-B}$ for which $f(\alpha) = \alpha$ and $f'(\alpha) = 1$ for the $F-B$ measure, and we obtain the entropic or information dimension of the measure to be $-\frac{1}{2}\frac{\lambda_1\log \lambda_1 + \lambda_2\log \lambda_2 +\dots}{\log(c2^{\lambda_1} 3^{\lambda_2} \dots)}$, with $\lambda_j = \frac{1}{2^j}$ and $c = \sqrt{\frac{\pi^2}{6}-1}$, which yields the value $0.87038$ in the interval defined by $0.870 \pm 0.0004$ (again Bak 1986), the universal constant associated with the dynamics of the Circle Map \textemdash which is why we conjecture that said dynamics inherits, via the $F-B$ structured staircase, this universal constant, which is an inherent property of the $F-B$ measure.
\newline\indent
Finally, by taking averages over the very different lengths of intervals in Nth $F-B$ partitions, as $N$ grows, we obtain a statistical $F-B$ contractor (Sec. 7) with which we can calculate the dimensionally \emph{Euclidean} (having only one contractor, all segments have equal Euclidean length at any Nth partition), statistically self-similiar, fractal version of the $F-B$ partition. Such process, briefly described in Sec. 7 yields again the universal value $\sim 0.870$ (Sec. 8), which is a second and much deeper connection between the two measures. 
\newline\indent
The definition of the $F-B$ measure on $I$ as constant over intervals in Nth partitions is not an arbitrary one: there is a non-Euclidean geometry on the upper half-plane, the $F-B$ partition is its inheritance on $I$. This geometry (we briefly comment on it in Sec. 9) has an associated regular tiling which partitions the real line in $F-B$ interpolations, and the location of the tiles approaching a real irrational number "\emph{i}" describes \textemdash by naked-eye direct observation\textemdash\space its decomposition in continued fractions, which yields, as we will see, the location of "\emph{i}" in the multifractal spectrum of the $F-B$ measure.
\newline\indent
NOTE: Sec. 6 can be by-passed by the reader: there are old and new results on the $(\alpha,f(\alpha))$ spectrum of the $F-B$ measure, and in Sec. 6 we "harmonize" the \textemdash only apparent\textemdash\space corresponding discrepancies. 
\newline\indent
NOTE: Although the Math level in this paper does not go beyond finding extremes of a function of several variables, the reader un-interested in long and tedious and tiresome estimates, approximations, and calculations can proceed to Sec. 2: Generalities and Notations, then go to Sec. 7 and therefrom to Sec. 9: Geometrical Considerations, Conclusions and Conjectures.
\newline\indent
IMPORTANT NOTE: By necessity we work, from Sec. 3 to Sec. 8, sometimes with approximations "$\cong$", sometimes with exact equalities "$=$", so the corresponding calculations yield estimates, and make no claim to be rigorous proofs of formal theorems. 
\section{Generalities and Notation}
With $p_j$ we will denote probabilities, $\lambda_j$ will be frequencies, $c_j$ contractors, $\mathcal{E}   $ will be a \textquotedblleft normalizing sum" (a different one for each normalizing process); $\Lambda$ and $\mu$ the coefficients of the Lagrange method of indeterminate coefficients for finding extremes of functions. With \textquotedblleft$i$" we will denote an irrational number in the unit interval,
$
i=\frac{1}{{a_1+\frac{1}{{a_2+...}}}}:= [a_1,a_2,\dots a_n, \dots ]
$
is its continued fraction expansion; $a_j$ the so-called partial quotient coefficients will be natural numbers; the rational number $[a_1 \dots a_n] := \frac{p_n}{q_n}$ is the nth rational approximant to $i$; $q_n$ the so called nth cumulant.
\newline
\indent The Farey-Brocot tree interpolates rational $\frac{a+a'}{b+b'}$ between rationals $\frac{a}{b}$ and $\frac{a'}{b'}$, starting from $\frac{0}{1}$ and $\frac{1}{1}$, the extremes of the unit interval. The first interpolation has, therefore, two segments $[\frac{0}{1},\frac{1}{2}]$ and $[\frac{1}{2},\frac{1}{1}],\dots$ the Nth step of interpolation partitions the unit segment in $2^N$ segments. Approximant $\frac{p_n}{q_n} = [a_1 \dots a_n]$ appears in the Nth step of the partition process, $N=a_1+\dots...+a_n;$ $i \in [\frac{p_n}{q_n}, \frac{p_{n+1}}{q_{n+1}}]$ $\forall n \in N$ (provided  that $\frac{p_n}{q_n} < \frac{p_{n+1}}{q_{n+1}}$), and the length of this segment is $\frac{1}{q_nq_{n+1}}$.
For a certain probability measure on the unit segment, let us consider a partition, $l_j$ the length of its segments, $p_j$ their probability measure. Let $q$ and $\tau$, real numbers, be connected through $\sum_j{\frac{{p_j}^q}{{l_j}^\tau}}=1$,  $\tau = \tau(q)$. Then, the so called thermodynamical formalism yields the multifractal spectrum $(\alpha,f(\alpha))$ of the probability measure in terms of lagrangian coordinates: $\alpha = \tau'(q);$ $f(\alpha)=\alpha q- \tau;$ $q=f'(\alpha); f''(\alpha)<0, $ and $q=1$ iff $\alpha=f(\alpha)$.
Let us recall that $f(\alpha)$ is, theoretically, the Hausdorff dimension of the subfractal $\Omega_\alpha$ which contains all elements with the same $\alpha$-concentration; and that the $\alpha$-concentration of a segment is the log-log version of the density: $\frac{\log p}{\log l},$ $l$ the length of the segment and $p$ its probability measure. Point $\alpha$-concentration is defined in the same way as point density: from $\alpha$-concentration of segments containing the point and a limiting process.
\section{The Euclidean Case: Equal Lengths}
\subsection{Two initial probabilities}
First we consider the unit segment as partitioned in two segments of equal Euclidean length when $N=1$,\dots\space and in $2^N$ Euclidean equal parts in step $N$. The first two segments have probability measures $p_1$ and $p_2$, $p_1+p_2=1$, so $p_j$, $j=1,2$ are contractions. The four segments of equal length in step $N=2$ have probability measures ${p_1}^2$, $p_1 p_2$, $p_2 p_1$ and ${p_2}^2$ respectively \dots and so on. In step $N$, we have $2^N$ segments, and their generic probability is ${p_1}^r {p_2}^{N-r}$, $r$ an integer, $0\leq r \leq N$. The number of segments with this probability is ${N \choose r}$. We will redo a calculation of the multifractal spectrum $(\alpha,f(\alpha))$ of this measure in terms of a key frequency $\lambda = \frac{r}{N}$ as internal coordinate. In step $N$ we have, for an \textquotedblleft r" segment, a concentration $\alpha = \frac{\log prob.}{\log length} = \frac{\log({p_1}^r {p_2}^{N-r})}{lg(\frac{1}{2^N})} = \frac{r \log p_1 + (N-r)\log p_2}{-N \log 2} = \frac{-1}{\log 2}(\frac{r}{N}\log p_1 + (1-\frac{r}{N})\log p_2) = \frac{-1}{\log 2}(\lambda\log p_1 + (1-\lambda)\log p_2)$. We can write then $\alpha(\lambda)$ for this magnitude. The number of such elements is ${N \choose r} = \frac{N!}{r!(N-r)!} \cong \frac{N^N}{r^r (N-r)^{N-r}}\frac{1}{e} = \frac{1}{e}\{\frac{1}{(\frac{r}{N})^{\frac{r}{N}}(1-\frac{r}{N})^{1-\frac{r}{N}}}\}^N = \frac{1}{e}(\frac{1}{\lambda^\lambda (1-\lambda)^{1-\lambda}})^N$. Therefore, in that step, and for such $\alpha$, we have $f(\alpha)=\frac{\log \frac{1}{e}(\frac{1}{\lambda^\lambda (1-\lambda)^{1-\lambda}})^N}{\log 2^N} = \frac{-1}{\log 2}\{\lambda\log\lambda + (1-\lambda)\log(1-\lambda)\}$ if $N$ tends to $\infty$. Our system now reads $\alpha = \frac{-1}{\log 2}(\lambda\log p_1+(1-\lambda)\log p_2) = \frac{-1}{\log 2}(\lambda\log p + (1-\lambda)\log(1-p))$; $p_1:=p$; $f(\alpha) = \frac{-1}{\log 2}(\lambda\log\lambda+(1-\lambda)\log(1-\lambda))$, from which it is obvious that $\alpha=f(\alpha)$ iff $\lambda=p$. Also $f'(\alpha) = \frac{df}{d\alpha}= \frac{df/d\lambda}{d\alpha/d\lambda}= \frac{\log\frac{\lambda}{1-\lambda}}{\log\frac{p}{1-p}}$ and again, $f'(\alpha) = 1$ iff $\lambda = p$ iff $\alpha = f(\alpha)$.  As variable $\lambda:0 \rightarrow 1$ varies, the $(\alpha,f(\alpha))$ graph is drawn. We call $\lambda$ an \textquotedblleft internal" coordinate, because in a certain step $N$, $\lambda = \frac{r}{N}$  tells us the value of $r$, the number of times in which $p$ appears, that is the proportion $\frac{r}{N}$ of all segments with measure ${p_1}^r {p_2}^{N-r} = p^r(1-p)^{N-r}$.
\indent
Henceforth, we will be interested in the subfractal for which $\alpha = f(\alpha)$ and $f'(\alpha)=1$, in all cases and in all measures considered. That we have, for this case, $f'(\alpha)=1$ iff $\alpha=f(\alpha)$ validates the thermodynamical formalism (see Sec. 2), which is not at all proved to yield the multifractal spectrum of an arbitrary measure, but which holds true for the Euclidean measure. In this case $\sum_{j} \frac{{p_j}^q}{{l_j}^\tau} = 1$ for $f'(\alpha)=q=1$ yields $\sum_{j} \frac{{p_j}}{{l_j}^\tau} = 1$  which, together with $\sum p_j = 1$, and the fact that partition $\{l_j\}$ is arbitrary, imply $\tau = 0$, which is another feature of the thermodynamical algorithm: $\alpha=f(\alpha)$ iff $q=f'(\alpha)=1$ iff $\tau(q)=0$.
\newline
\indent
Next, let us arrive at the expressions for $\alpha,f(\alpha)$, and $f'(\alpha)$ through the lagrangian coordinates in the thermodynamical algorithm, and compare said expessions with those above, with $\lambda : 0 \rightarrow 1$ as the internal coordinate. From  $\sum_{j} \frac{{p_j}^q}{{l_j}^\tau} = 1$, and $l_j \equiv \frac{1}{2^N}$ $\forall j$, in step $N$, we have 
\[\tau \cong \frac{\log \sum_{j} {p_j}^q}{\log\frac{1}{2^N}} =
-\frac{1}{N\log 2}\log\sum_{r=0}^N(p^r(1-p)^{N-r})^q {N \choose r} =\]\[ 
\frac{-1}{N\log 2}\log\sum_{r=0}^N[p^q]^r[(1-p)^q]^{N-r} {N \choose r} = 
\frac{-1}{N\log 2}\log(p^q+(1-p)^q)^N=\]\[
\frac{-1}{\log 2}\log(p^q+(1-p)^q).
\]
\indent
So \[ \alpha = \tau'(q) = -\frac{1}{\log 2}\frac{1}{p^q+(1-p)^q} \{ p^q\log p + (1-p)^q\log(1-p) \}= \]
\[
 \frac{-1}{\log 2}\{\frac{p^q}{p^q+(1-p)^q}\log p + \frac{(1-p)^q}{p^q+(1-p)^q}\log(1-p)\}; 
\]
a comparison with $\alpha=\frac{-1}{\log 2}\{\lambda\log p + (1-\lambda)\log(1-p)\}$ yields 
\begin{equation} \label{eq:eq1}
\lambda = \frac{p^q}{p^q+(1-p)^q}.
\end{equation}
\indent
Next, following the algorithm, we have $f(\alpha) = \alpha q - \tau = $
\[
	= q\frac{-1}{\log 2}\{\frac{p^q}{p^q+(1-p)^q}\log p + \frac{(1-p)^q}{\mathcal{E} }\log(1-p) \} - \frac{-1}{\log 2}\log \mathcal{E} = 
\]
\[
	= \frac{-1}{\log 2}\{\frac{p^q}{\mathcal{E}}\log p^q+ \frac{(1-p)^q}{\mathcal{E}}\log(1-p)^q - (\log \mathcal{E})[\frac{p^q}{\mathcal{E}}+\frac{(1-p)^q}{\mathcal{E}}]\} = 
\]
\[
	= \frac{-1}{\log 2}\{\frac{p^q}{\mathcal{E}}\log\frac{p^q}{\mathcal{E}}+\frac{(1-p)^q}{\mathcal{E}}\log\frac{(1-p)^q}{\mathcal{E}}\}
\]
and again, if we compare with the value $\frac{-1}{\log 2}\{ \lambda\log\lambda+(1-\lambda)\log(1-\lambda)\}$ we obtain $\lambda = \frac{p^q}{\mathcal{E}}$. Now, $f'(\alpha) = \frac{\log\frac{\lambda}{1-\lambda}}{\log\frac{p}{1-p}}$, with this value of $\lambda$, and with $\frac{\lambda}{1-\lambda}=\frac{p^q}{(1-p)^q}$, since $\mathcal{E}$ cancels, becomes 
\begin{equation} \label{eq:eq2}
f'(\alpha) = \frac{\log\frac{p^q}{(1-p)^q}}{\log\frac{p}{1-p}} = q
\end{equation}
indeed.
\newline
\indent
This will be the procedure for the next sections: to express $\alpha$ and $f(\alpha)$ in terms of contractors and key frequencies, to find an expression for these frequencies in terms of the thermodynamical parameters $q$ and $\tau$, and to find the frequencies for which $\alpha = f(\alpha), f'(\alpha)=1,$ and $\tau = 0$.

\subsection{A finite number of initial probabilities}
\indent
Consider, next, the unit segment as partitioned in $n_0$ segments of equal  Euclidean length when $N=1$, \dots and in ${n_0}^N$ equal parts in step $N$. The first $n_0$ segments for $N=1$ have probability measures $p_1,\dots,p_{n_0}$; $p_1+\dots +p_{n_0}=1$; $p_j$ contractors. This case is \emph{quite} different from that in which $n_0=2$: the frequencies of $p_1$ and $p_2=1-p_1$ were $\lambda_1$ and $\lambda_2 = 1 - \lambda_1,$ so there was a coordinate $\lambda=0 \rightarrow 1$, a \textquotedblleft natural" or \textquotedblleft internal" coordinate in charge of producing the spectrum. We cannot have that convenience here, for the frequencies of the $p_j$, the $\lambda_j$, will be $\lambda_1 \dots \lambda_{n_0}$, $\sum\lambda_j=1$, so we have many independent coordinates.
\newline
\indent
Let $N$ be the step, $r_1\dots r_{n_0}$ a particular choice of integers, $0 \leq r_j \leq N$; $\sum r_j = N,$ we consider segments of length $\frac{1}{n_0^N}$ with measure $p_1^{r_1}\dots p_{n_0}^{r_{n_0}}$. We proceed as in the previous section: $\lambda_j=\frac{r_j}{N}$, $\sum_j\lambda_j=1,$ $ const\cdot\frac{1}{(\lambda_1^{\lambda_1}\dots)^N}$ the number of such segments with the $\{r_j\}$ or the $\{\lambda_j\}$ particular choice. The $\alpha$-concentration and the $f(\alpha)$ corresponding to such a set $\{\lambda_j\}$ are: $\alpha(\lambda_1 \dots) = \frac{\lambda_1\log p_1+\dots}{-\log n_0}$; $f(\alpha(\lambda_1\dots))= \frac{\lambda_1\log \lambda_1+\dots}{-\log n_0}$, proceeding as in the previous section. But the difference with last section arises now: for a fixed value of $\alpha$ we are interested in \emph{all} choices of $\{\lambda_j\}$ which fulfill $\alpha(\lambda_1\dots\lambda_j\dots)=\alpha$. And the dimension $f(\alpha)$ of this subfractal will be the maximum value of $f(\alpha(\lambda_1\dots))$ which fulfills $\alpha(\lambda_1\dots)= \alpha$ and $\sum\lambda_j=1$. Therefore, we have to extremize $-\frac{1}{\log n_0}(\lambda_1\log\lambda_1+\dots)-\Lambda\frac{-1}{\log n_0}(\lambda_1\log p_1+\dots-\alpha)+\mu(\lambda_1+\dots-1)$, with $\lambda_j$ as variable. The corresponding calculations are shown in the App. to Sec. 3.2; the result: 
\begin{equation}  \label{eq:eq3_1}
\lambda_j= \frac{p_j^\Lambda}{\mathcal{E}} \qquad \forall j
\end{equation}
with 
\begin{equation} \label{eq:eq4_1}
\Lambda=\Lambda(\alpha), \textrm{ or } \alpha=\alpha(\Lambda)
\end{equation}
which yields $f'(\alpha) = \Lambda$.
\newline \indent
So, the lagrangian indeterminate coefficient $\Lambda$ fulfills four roles: (1) it is the lagrangian coefficient linking $f(\alpha)$ with $\alpha$; (2) it is the exponent, in Eq. \ref{eq:eq3}, of $p_j$, which, normalized, determines $\lambda_j$; (3) it gives $\alpha$ from Eq. \ref{eq:eq4}, and (4) it is $f'(\alpha)=q$, as we have just seen. Notice the similitude between these results and Eqs. \ref{eq:eq1}, \ref{eq:eq2}.
	From the values of $\alpha$ and $f(\alpha)$ obtained in the Appendix to Section 3.2 (see  Eq. \ref{eq:eq5}), we can see that $\alpha=f(\alpha)$ iff $\lambda_j=p_j$, which happens iff $\Lambda=1$, i.e. if $f'(\alpha)=1$. All of which implies $\tau=0$.
\section{The Euclidean Case: Equal Probabilities}
In this case the lengths of all $n_0^N$ segments in the partition of the unit segment corresponding to the Nth stage or step of the construction of the multifractal are given by contractors $c_1\dots c_{n_0}$, a natural extension of the case in Sec. 3. All of the $n_0^N$ probabilities are equal. With $\lambda_j$ as before, the generic length of such a segment is $(c_1^{\lambda_1}\dots c_{n_0}^{\lambda_{n_0}})^N$. Proceeding as in Sec. 3. we have to extremize the function $\frac{\lambda_1\log\lambda_1+\dots}{\lambda_1\log c_1+\dots} - \Lambda(\frac{-\log n_o}{\lambda_1\log c_1+\dots}-\alpha)+ \mu(\lambda_1+\dots - 1)$ with $\lambda_j$ as variable. The corresponding calculations are shown in Appendix 1 to Sec. 4; the result: 
\begin{equation} \label{eq:eq6}
	\lambda_j = \frac{c_j^\Xi}{\mathcal{E}}
\end{equation}
which bears a resemblance to Eq. \ref{eq:eq3_1}; here $\Xi$ is an exponent of the contractor $c_j$, and
\begin{equation} \label{eq:eq7_1}
\nonumber
f'(\alpha)= \frac{\log \mathcal{E}}{\log n_0}
\end{equation}
We want to interpret Eq. \ref{eq:eq7_1}, since it does not look anything like. 
\newline
\indent
Let $q = f'(\alpha),\tau(q),\alpha,f(\alpha),$ be the thermodynamical magnitudes involved in the process described in Sec. 3: equal lengths and different probabilities. In Sec. 4 we are reversing the process, exchanging the role of lengths and probabilities: equal probabilities and different lengths, which has been termed \textquotedblleft the inverse process". Notice that $\sum_j p_j=1$ and $\sum_j c_j=1$ make this inversion totally plausible. Let $\bar{f},\bar{\alpha},\bar{q},\bar{\tau}$, be the new thermodynamical parameters. 
\newline\indent
In Appendix 2 to Section 4 we deduce the relationships between "old" $\alpha, f, q = f'(\alpha), \tau$ and the new $\bar{f},\bar{\alpha}, \bar{q}, \bar{\tau}$:
\begin{equation} \label{eq:eq8}
\bar{\alpha}=\frac{1}{\alpha},
\end{equation}
\begin{equation} \label{eq:eq9}
-\tau(q)=\bar{q} \qquad \forall q
\end{equation} 
\begin{equation} \label{eq:eq10}
-\bar{\tau}(-\tau(q))= q \qquad \forall q \qquad \mathrm{or} -\tau(-\bar{\tau}(\bar{q})) = \bar{q}  \qquad \forall \bar{q} 
\end{equation}
which yield 
\begin{equation} \label{eq:eq11}
\bar{q} = \frac{d\bar{f}}{d\bar{\alpha}} = \frac{\log \mathcal{E}}{\log n_0} = - \tau(\Xi) 
\end{equation}
and $\Xi = - \bar{\tau}(\Lambda)$, from which $\bar{q} = -\tau(-\bar{\tau}(\Lambda)) = \Lambda$ or $\bar{q} = \Lambda$ in Sec. 4, as $q = \Lambda$ in Sec. 3.
\newline
\indent
The expression $\Xi = -\bar{\tau}(\Lambda)$ above becomes $\Xi = -\bar{\tau}(\bar{q})$, which is $q$, by Eq. \ref{eq:eq9} \textemdash that is, the \textquotedblleft old" q. Hence, cases in Secs. 3 and 4 have an analogy: the lagrangian coefficient $\Lambda$ is $f'$ in both spectra, and a difference: the exponent of the contractor giving the critical $\lambda$ is, in the first case, $\Lambda = q $ and, in the second case, the $q$ of the \emph{inverse} problem; i.e. the $\Lambda$ of the \emph{inverse} problem.
\newline
\indent
Notice that condition $\alpha = f(\alpha)$ (in Sec. 3/Sec. 4 notation we should write $\bar{\alpha} = \bar{f}(\bar{\alpha})$) is fulfilled for $\lambda_j = \frac{1}{n_0}$, for then $\lambda_1\log\lambda_1+\dots=\frac{1}{n_0}\log\frac{1}{n_0}+\dots = n_0\frac{1}{n_0}(-\log n_0) = -\log n_0$, the numerator of $\alpha$. But $\lambda_j = \frac{1}{n_0}$, which from Eq. \ref{eq:eq6} means $\Xi = 0$, i.e. $ -\bar{\tau}(\Lambda) = 0$, which means $\Lambda = f'(\alpha) = 1$.
\newline
\indent
Again $f(\alpha)=\alpha$, $f'(\alpha)=1$, $\tau=0$ are simultaneous conditions in order to characterize the subfractal which concentrates the measure, $f(\alpha)$ being the entropic or information dimension.
\section{The $F-B$ Case}
\subsection{Equal probabilities and different lengths}

The treatment of the thermodynamical multifractal spectra in the Euclidean case, expressing key parameters in terms of contractors and their frequencies in Secs. 3 and 4, permits \textemdash we trust\textemdash\space extending such results and reasonings to the case of the $F-B$ measure on the unit interval given by the Farey Brocot F-B partition tree. As in Sec. 4 we deal with equal probabilities and different lengths.
\newline 
\indent
The Nth step or stage of the F-B interpolation gives a partition of the unit segment in $2^N$ smaller segments of equal $\frac{1}{2^N}$ probability. Let $a_1\dots a_n$ be positive integers such that $a_1+\dots+a_n=N$. There is a segment in that step of length $\frac{1}{q_{n-1}q_n}$, where $[a_1 \dots a_n]= \frac{p_n}{q_n}$ (see Sec. 2). This segment contains all irrationals of the form $i=[a_1,\dots,a_n,$ etc. $]$, where \textquotedblleft etc." is \emph{any} sequence of natural numbers $a_{n+1},\dots,a_j,\dots$.
\newline
\indent
We want to interpret nested segments of length $\frac{1}{q_nq_{n-1}},\frac{1}{q_{n+1}q_n},\dots $ in terms of contractors.

\subsection{The $F-B$ contractors}
First we observe that lengths of nested intervals diminish like $\frac{1}{q_n^2}$: since $q_{n+1} = a_{n+1}q_n+q_{n-1}$, the $q_n$ grow with $n$, hence $\frac{1}{q_{n+1}^2}< \frac{1}{q_nq_{n+1}}<\frac{1}{q_n^2}$. Therefore, we will estimate lengths $\frac{1}{q_nq_{n+1}}$, $ q_n=q_n(a_1,\dots,a_n)$ by $\frac{1}{q_n^2}$ in step $N=a_1+\dots+a_n$. Now, 
\begin{equation} \label{eq:eq12}
a_{n+1}q_n<q_{n+1}=a_{n+1}q_n+q_{n-1}<a_{n+1}q_n+q_n= (a_{n+1}+1)q_n.
\end{equation}
So, the contractor that shrinks length $\frac{1}{q_n^2}$ into the smaller one $\frac{1}{q_{n+1}^2}$ is a number somewhere between $\frac{1}{a_{n+1}^2}$ and $\frac{1}{(a_{n+1}+1)^2}$. Now, a moment of reflection observing Eq. \ref{eq:eq12} shows that $q_{n+1}$ is much nearer $(a_{n+1}+1)q_n$ than $a_{n+1}q_n$: $a_{n+1}q_n$ is far smaller than $q_{n+1}$ because an exponential $(q_{n-1})$ is missing, whereas by replacing $q_{n-1}$ by $q_n$ (in the RHS of Eq. \ref{eq:eq12}) we just replace one exponential by another which could be connected to the first one by a reasonable coefficient.
\newline
\indent
Therefore, a certain contraction $c(a_{n+1}+1)$ of $a_{n+1}+1$ will yield $q_{n+1}$ from $q_n$, via Eq. \ref{eq:eq12}, hence $\frac{1}{q_{n+1}^2}$ from $\frac{1}{q_n^2}$.
\newline
\indent
Let us consider $q_n=q_n(a_1\dots a_n)$. Magnitude $q_n$ is obtained from $q_{n-1}$ via $c(a_n+1)$, we still do not know the value of c. Iterating this process we have $q_n(a_1\dots a_n)$ given by $c(a_1+1)c(a_2+1)\dots c(a_n+1)$. The $n$ integers $a_1\dots a_n$ vary \textemdash say\textemdash\space between $1$ \& $k \in  \mathds{N}$. Let $r_j$ be the number of times for which the $a$'s are equal to $j:1 \rightarrow k$, $\sum_j r_j = n$. Then $q_n$ is given by $[c(1+1)]^{r_1} \dots [c(k+1)]^{r_k} = c^{r_1+\dots+r_k}2^{r_1}\dots(k+1)^{r_k} = c^n[2^{\lambda_1}\dots (k+1)^{\lambda_k}]^n = [c2^{\lambda_1}\dots (k+1)^{\lambda_k}]^n$. Here $k$ is, simply, the largest of the integers $a_1 \dots a_n$, $\lambda_j=\frac{r_j}{n}$, and $\lambda_1+\dots=1.$
\newline
 {\bf We rewind}: segment of length $\frac{1}{q_{n+1}^2}$ is obtained from that of length $\frac{1}{q_n^2}$ through a contraction $\frac{1}{[c(a_{n+1}+1)]^2}$, $a_{n+1}$ an integer. The estimate 
\begin{equation} \label{eq:eq13}
q_n \cong [c 2^{\lambda_1}\dots(k+1)^{\lambda_k}]^n
\end{equation}
above, $c$ an appropriate constant, is a simplified version of the Besicovitch formula [Good, 1941], which we have already used elsewhere [Piacquadio, 2004]. We are in the F-B step $N=a_1+\dots+a_n$, in a segment of length estimated by $\frac{1}{q_n^2} \cong \frac{1}{[c 2^{\lambda_1}\dots(k+1)^{\lambda_k} ]^{2n}}$, $k$ simply the largest value of the $a_j$, $j:1 \rightarrow n$. The probability ($F-B$) measure of such segment is $\frac{1}{2^N} = \frac{1}{2^{\sum_{j=1}^n a_j}}$.

We need, now, to estimate constant $c \in (0,1)$. This we do in the Appendix to Sec. 5.2, by estimating the Hausdorff dimension $d_H$ of $E_k := \{ i = [a_1\dots a_j \dots ] / a_j \leq k $ $\forall j \}, k \in \mathds{N}$, $\lambda_j$ the frequency in which the $a$'s are equal to $j$. We use a result of Jarnik [1928; 1929] who proved $1-d_H(E_k) = O(\frac{1}{k})$, and obtain the $\lambda_j$ responsible for the dimension: 
\begin{equation}
\label{eq:13_1}
\lambda_j = \frac{(j+1)^a}{\mathcal{E}}
\end{equation}
with 
\begin{equation}
\label{eq:13_1}
a = -2 d_H(E_k)
\end{equation}
$a \cong -2$ as $k$ grows, which implies $c = \sqrt{c_\pi}$; $c_\pi = \frac{\pi^2}{6} - 1$.
\newline
\indent
From our \textquotedblleft We rewind" note above, we have the generic value of the $F-B$ contractors: $ \frac{1}{[\sqrt{c_\pi}(a_{n+1}+1)]^2}$ and, since $a_{n+1}$ is any integer $j$, $\frac{1}{c_\pi(j+1)^2}$ is the generic contractor, $j \in \mathds{N}$. The main difference with the Euclidean case is that we have an infinity of contractors now. 
\subsection{A first estimate of the $(\alpha,f(\alpha))$ spectrum for the $F-B$ measure}
From the preceding section the probability of segment with length $1/[c 2^{\lambda_1}\dots(k+1)^{\lambda_k}]^{2n}$ is $1/2^N = 1/2^{\sum_1^n a_j}$; $k$ the largest of the $a$'s. The $\alpha$-concentration of this segment is, then,
\begin{equation} \label{eq:eq18}
\alpha = \frac{-\log 2}{-2n}\frac{\sum_{j=1}^n a_j}{\log c+\lambda_1\log 2+\dots}=\frac{\log 2}{2}\frac{(\sum_1^n a_j)/n}{\log c+\lambda_1\log 2+\dots} := \frac{\log 2}{2}\frac{ m }{\log c+\lambda_1\log 2+\dots}
\end{equation}
Integer $n$ \textquotedblleft disappears" in the average value $m$ of the $a$'s, whereas $k$ will become quite relevant. 
\newline
\indent
Let us consider, in $E_k$, the set $S_m$ of elements with average of the $a$'s no larger than $m$ \textemdash technically, it should be $ \limsup_n(\sum_1^n a_j)/n \leq m$, but the essential idea is to control the average of the $a$'s. A choice of $\lambda$'s: $\{\lambda_1\dots \lambda_k\}$ will label different subsets of $S_m$. As we saw above, the subset of largest dimension corresponds to the label $\lambda_j \cong \frac{(j+1)^{-2}}{c_\pi}$, and we have to add the extra condition on the size of the average of the $a$'s. This particular choice of $\lambda$'s is both responsible for the dimension of $S_m$ and, therefore, for the value of $\alpha$ associated with it, which, from Eq. \ref{eq:eq18} becomes $\alpha = \frac{m}{K}, K$ the denominator in Eq. \ref{eq:eq18} for these particular $\lambda$'s. 
\newline
\indent
Now, let $r_j$ be, as before, the number of $a$'s equal to $j$, $\lambda_j = \frac{r_j}{n}$, then $\frac{\sum a_j}{n} = 1\lambda_1 + 2\lambda_2+\dots+ k\lambda_k = \frac{1}{c_\pi}\sum_1^k\frac{j}{(j+1)^2} \cong \frac{\log k}{c_\pi} \leq m$ \dots so $k$ cannot be larger than $k_m = e^{c_\pi m} = e^{c_\pi\alpha K} := e^{B\alpha}$, \textquotedblleft $B$" a constant. Applying the already quoted result by Jarnik, refined by Hensley [1996], we have $f(\alpha) = 1 - \frac{\mathrm{const}}{k_m} := 1 - \frac{A}{e^{B\alpha}}$, but only if $m$ \textemdash and therefore $\alpha$\textemdash \space is not too small.
\newline
\indent
The result is partially hinted at by Cesaratto and Piacquadio [1998],  Piacquadio and Cesaratto [2001], and Piacquadio [2004], and in Piacquadio [2004] it is empirically shown to be computationally correct within relatively small percentage errors. 
\newline
{\bf Note:} The value of 
\begin{equation} \label{eq:eq19}
 \lambda_j \cong \frac{(j+1)^{-2}}{c_\pi} = 
\frac{(j+1)^{-2}}{\mathcal{E}}
\end{equation}
just quoted, responsible for $f(\alpha)$ in the $F-B$ case, when $\alpha$ is not small, can be refined a bit. Let us remember that the exponent \textquotedblleft -2" comes from Eq. \ref{eq:eq17}: the exponent is $a = -2 d_H(E_k) \cong -2$ if $k$ (and $k_m$, and $m$, and $\alpha$) is large. Remembering also (see the end of Sec. 5.2) that $\frac{1}{c_\pi(j+1)^2}$ is the generic $F-B$ contractor, and that $d_H(E_k)$ with $E_k$ restricted by $(\sum_j a_j)/n \leq m$ becomes $f(\alpha)$ in this section, we finally have 
\begin{equation} \label{eq:eq20}
\lambda_j = \frac{(j+1)^{-2f(\alpha)}}{\mathcal{E}}
\end{equation}
which can be written as $\frac{[\frac{1}{c_\pi(j+1)^2}]^{f(\alpha)}}{\mathcal{E}}$,  where the last $\mathcal{E}$ normalizes the introduced factor $(\frac{1}{c_\pi})^{f(\alpha)}$. So we have $\lambda_j = \frac{[ \mathrm{j^{th}contractor} ]^{f(\alpha)}}{\mathcal{E}}$. Notice that this value has much in common with the critical $\lambda$'s for the Euclidean case: for equal lengths we had 
\begin{equation} \label{eq:eq21}
\lambda_j = \frac{[ \mathrm{j^{th}contractor} ]^{\mathrm{exponent}}}{\mathcal{E}},
\end{equation}
as for equal probabilities, and again for the $F-B$ measure. The difference is in the value of the exponent: $f'(\alpha)$ for equal lengths in the Euclidean case, $\overline{f}'(\alpha)$ for equal probabilities, same case, $f(\alpha)$ for the $F-B$ one\dots We will return to these \emph{apparent} differences later on, in Sec. 6. For now, we want to stress the universal character of Eq. \ref{eq:eq21}, where Euclidean and $F-B$ measures intersect. 
\subsection{A better expression for $(\alpha,f(\alpha))$}
We want now a more accurate expression for $\alpha$ and $f(\alpha)$ for the $F-B$ measure. With the same notation as in Sec 5.3, we want to extremize $\frac{\log(\frac{1}{\lambda_1^{\lambda_1}\dots \lambda_l^{\lambda_k}})^n}{ \log \{\{([c 2^{\lambda_1}\dots (k+1)^{\lambda_k}]^n)^2\}\}} - \Lambda(\frac{\log \frac{1}{2^{\sum a_j}}}{\log(1/\{\{\}\})} - \alpha )$, with condition $\sum \lambda_j = 1$, $\lambda_j$ the variable. This is done in the Appendix to Section 5.4. The critical $\lambda_j$ are 
\begin{equation} \label{eq:eq22_1}
\lambda_j = \frac{(j+1)^{2\tau(\Lambda)}/2^{\Lambda(j-1)}}{\mathcal{E}}.
\end{equation}
Now this value of $\lambda_j$ \emph{seems} to be very different from those obtained in Secs. 3 and 4, and from those in Eqs. 17, 18, and those from previous work [Piacquadio and Cesaratto, 2001]. We will show the corresponding connections in Sec. 6.
\subsection{The information dimension for the $F-B$ measure}
In this section, we will find the value of $\alpha$ for which $f(\alpha)=\alpha$ and $f'(\alpha)=1$, showing that this entropic or information dimension is the universal value $0.870 \pm 0.0004$ found by Bak and others [Bak, 1986 and references] to be the approximated box dimension of the fractal underlying the Cantor staircase for the circle map, in frontier with Chaos. 
\newline
\indent
Equating $f(\alpha)$ and $\alpha$ we obtain $-\frac{1}{2}\frac{\lambda_1\log \lambda_1+\dots}{\log(c 2^{\lambda_1}\dots)} = \frac{\log 2}{2} \frac{\sum_j j\lambda_j}{\log(c 2^{\lambda_1}\dots)}$, which implies $\sum_j(\lambda_j\log\lambda_j +(\log 2)j\lambda_j) = 0$ or $\sum_j\lambda_j(\log\lambda_j + \log 2^j) = \sum_j\lambda_j\log(\lambda_j 2^j) = 0$, so, if we write $\lambda_j = \frac{1}{2^j}$ we have $\sum_j\lambda_j=1$ and $f(\alpha) = \alpha$. 
\newline
\indent
With this particular value of $\lambda_j$, Eq. \ref{eq:eq22} now reads $\frac{1}{2^j} = \frac{(j+1)^{2\tau(\Lambda)}/2^{\Lambda(j-1)}}{\mathcal{E}}$, which can be rewritten (with $\mathcal{E}$ always the corresponding normalizing sum):
\begin{equation} \label{eq:eq23}
\frac{1}{2^j} = \frac{(j+1)^{2\tau(\Lambda)}}{2^{\Lambda j} \mathcal{E}}
\end{equation}
or $\frac{2^{\Lambda j}}{2^j} = 2^{(\Lambda-1)j} = \mathrm{const.}(j+1)^{2(\Lambda\alpha-f(\alpha))} = \mathrm{const}(j+1)^{2f(\alpha)(\Lambda-1)}$, since $\alpha = f(\alpha)$. So we have
\begin{equation} \label{eq:eq24}
2^{(\Lambda-1)j} = \mathrm{const}(j+1)^{2f(\alpha)(\Lambda-1)},
\end{equation}
with $2f(\alpha)$ a number strictly between $0$ and $2$. Now, if $\Lambda \neq 1$ we obtain from Eq. \ref{eq:eq24}: $2^j = \mathrm{const}(j+1)^{2f(\alpha)}$, obviously an absurdum as $j$ grows, so we confirm that $\alpha = f(\alpha)$ implies $\Lambda = 1$. On the other hand, let us assume that $\Lambda = 1$ above, in Eq. \ref{eq:eq23}. We are left with $\frac{1}{2^j}= \mathrm{const}\frac{(j+1)^{2(\alpha - f(\alpha))}}{2^j}$, which implies that $(j+1)^{2(\alpha-f(\alpha))}$ is a constant $\forall j$, an absurdum unless $\alpha = f(\alpha)$: so $\alpha= f(\alpha)$ iff $\Lambda=1$. But $\Lambda=1$ and $\alpha=f(\alpha)$ mean $\tau=0$, which seems to be in agreement with the Euclidean cases as \emph{the} condition that characterizes the concentration of the measure i.e. the information entropic dimension. 
\newline
\indent
For this case, in which $\lambda_j=\frac{1}{2^j}$, we have the corresponding $f(\alpha) = -\frac{1}{2}\frac{\lambda_1\log\lambda_1+\dots}{\log(c 2^{\lambda_1} 3^{\lambda_2}\dots)}$ to be $0.87038\dots$, the Hausdorff dimension of the subfractal which concentrates the $F-B$ measure. Notice that this number lies in the interval $0.870\pm 0.0004$ quoted above. The more restricted interval $0.870\pm 0.00037$ [Weisstein, 2005] for the box dimension of the fractal associated with the Circle map staircase, would differ from $0.87038\dots$ in one unit in the 5th decimal, an error that arises from the use of the simplified Besicovitch approximation \textemdash bound to be \textquotedblleft very good" indeed, according to Good [1941]\textemdash\space which does not take into account the \emph{order} in which the partial quotient coefficients $a_j$ appear in the cumulant $q_n=q_n(a_1\,\dots,a_n)$, but \emph{only} their \emph{values}.
\newline
\indent
{\bf Observations.} The formula for the key $\lambda_j$'s shown in Eq. \ref{eq:eq22} is much more complex than those for the Euclidean cases. An adaptation of the reasoning in Sec. 4, in order to prove that $\Lambda=f'(\alpha)$ in the $F-B$ case has been, so far, elusive. That is why we showed in some detail that, at least in the case of the subfractal that concentrates the $F-B$ measure, $\Lambda$ does act as $f'(\alpha)$. These efforts are necessary, when we recall that the validity of the thermodynamical formalism has been proved only for the Euclidean measures [Cawley \& Mauldin, 1992; Riedi \& Mandelbrot, 1997; 1998] and only semicomputationally for the $F-B$ measure.[Piacquadio \& Cesaratto, 2001]
\newline
There are old and new results on the $(\alpha,f(\alpha))$ spectrum of the $F-B$ measure, and in the next section we \emph{harmonize} the \textemdash only apparent\textemdash\space corresponding discrepancies \textemdash not all details included, for obvious limitations of scope and space, some fine brushings are left to the reader. The reader only interested in following the thread of the argument on $d \sim 0.870$ may skip Sec. 6.
\section{Relating the Key $\lambda_j$'s}
We seem to have two \textemdash apparently\textemdash\space very different expressions for the key $\lambda_j$'s in the case of the $F-B$ measure, which are, in turn, quite different from the key $\lambda_j$'s corresponding to the Euclidean case. Let us study these apparent discrepancies.
\newline
\indent
For the $F-B$ case, the value of $\lambda_j$ from Eq. \ref{eq:eq22} is $\lambda_j = \frac{(\frac{1}{(j+1)^2})^{-\tau}/2^{(j-1)\Lambda}}{\mathcal{E}} =  \frac{(\frac{1}{(j+1)^2})^{-\tau}/2^{j\Lambda}}{\mathcal{E}}$, and we want to connect this result with the value 
\begin{equation} \label{eq:eq25}
\lambda_j \cong \frac{1}{(j+1)^2}\frac{1}{c_\pi} ,
\end{equation}
$j:1 \rightarrow k_m = (e^{c_\pi})^m := e^{B\alpha}$, since average $m$ of the $a_j$ is proportional to $\alpha$, all according to Sec 5.3 ; $f(\alpha) \cong 1 - \frac{\mathrm{const}}{k_m} := 1 - \frac{A}{e^{B\alpha}}$, $A$ and $B$ positive constants, $B>1$ \textemdash and let us recall that this result was valid when $m$, and therefore $\alpha$, was not small.
\newline
\indent
Let us continue to assume that $\Lambda$ is the derivative of $f(\alpha$). Then $\Lambda = \frac{AB}{e^{B\alpha}} = \frac{\mathrm{const}}{k_m}$. If we recall that $j \leq k_m$ we have $\Lambda j \leq \frac{\mathrm{const}}{k_m}k_m$, so the value of $\Lambda j$, the exponent of $2^{\Lambda j}$ above, is \emph{bounded}, \dots so $\lambda_j$ from Eq. \ref{eq:eq22} is, essentially, $(\frac{1}{(j+1)^2})^{-\tau}$, normalized.
\newline
\indent
Now, let us have a closer look at the other expression (Eq. \ref{eq:eq25}) for the key frequency: $\lambda_j = \frac{1}{(j+1)^2}\frac{1}{c_\pi} = \frac{\frac{1}{(j+1)^2}}{\mathcal{E}}$, the approximant of $\frac{(\frac{1}{(j+1)^2})^{f(\alpha)}}{\mathcal{E}} = \frac{(j^{th}\mathrm{contractor})^{f(\alpha)}}{\mathcal{E}}$ according to Eq. \ref{eq:eq17}. The exponent $-\tau$ in the expression above, \textquotedblleft $(\frac{1}{(j+1)^2})^{-\tau}$, normalized", is $-\tau = f(\alpha) - \Lambda\alpha$, so Eq. \ref{eq:eq22} would be, essentially, $ (\frac{1}{(j+1)^2})^{f(\alpha)}(\frac{1}{(j+1)^2})^{-\Lambda\alpha}$, normalized; $j:1 \rightarrow k_m$. We want to analyze, therefore, the behaviour of the discrepancy between expressions \ref{eq:eq20} and \ref{eq:eq25}, i.e. $(\frac{1}{(j+1)^2})^{-\Lambda\alpha} = ((j+1)^{\Lambda\alpha})^2$, $j:1 \rightarrow k_m = e^{B\alpha}$ and $\Lambda = \frac{\mathrm{const}}{k_m} = \frac{\mathrm{const}}{e^{B\alpha}}, \alpha $ being proportional to $m$. So $(\frac{1}{(j+1)^2})^{-\Lambda\alpha} = (j+1)^{2\Lambda\alpha} \leq (k_m+1)^{2\Lambda\alpha} \cong (e^{B\alpha})^{\frac{\mathrm{const}}{e^{B\alpha}}\alpha} = e^{\frac{\alpha^2}{e^{B\alpha}}\mathrm{const}} \approx 1$. If $j$ does not grow, still the exponent $\Lambda\alpha$ tends to zero and, again, $(\frac{1}{(j+1)^2})^{-\Lambda\alpha} \approx 1$. So both expressions of the key $\lambda_j$'s are very much like $(\frac{1}{(j+1)^2})^{-\tau}$, normalized. Finally, if we recall that const.$\frac{1}{(j+1)^2}$ is the generic contractor in the $F-B$ construction of the Farey tree, then we have the key $\lambda_j$ given by $\frac{(j^{th}\mathrm{contractor})^{-\tau}}{\mathcal{E}}$, which is, exactly, the value for the key $\lambda_j$ in the Euclidean case. 
\section{A Statistical Version of the Farey Tree}
By connecting the cases where a segment is measured with a common Euclidean ruler, or by the $F-B$ probability $1/2^N$, we tried, so far, to establish a connection between Euclidean and $F-B$ measures, by means of their corresponding multifractal analysis. The differences between the two measures are considered to be deep and are briefly pointed at in Sec. 9. Yet, the thermodynamical algorithm \textemdash the multifractal spectrum\textemdash\space reveals, on a closer look, their inner links. We propose to deepen these links.
\newline
\indent
Let us suppose we are studying, empirically, the geometry of a fractal in a unit segment given by, say, a certain dynamical system, so we know the step $N$ in which we are. Further, let us suppose that the fractal is \textemdash once constructed, as $N$ grows\textemdash\space a ternary like that of Cantor, a typical self-similar "Euclidean" case in the sense described above. The subdivision of segments seems to correspond, empirically, to a left-right process, so we know that in step $N$ we have a list of $2^N$ segments. Their length seems to diminish exponentially, like $\frac{1}{A^N}, A>1$, but we are not sure of the value of $A$. We are not so much interested in the value of $A$, but on that of $\log A$, for we know that $\frac{\log 2}{\log A}$ would be the dimension that we are trying to estimate. In order to estimate $\log A$ (if we are in the ternary of Cantor, $\log A$ should be $\log 3$, but we are measuring experimentally) we take all $ 2^N$ segments in the Nth step, we take their reciprocals (so we would have $2^N$ segments of length $A^N$, roughly), we take their logarithms, we divide said logarithms by $N$\dots\space  and we take the average of all these values, for as large a value of $N$ as we can handle. That should give us a stable value converging to $\log A$, $\log 3$ if we were in the ternary of Cantor. 
\newline 
\indent
We propose to do such a calculation for the $2^N$ intervals in the Farey tree partition: we will take their Euclidean lengths, take their reciprocals, take their logarithms, divide them by $N$, and average all these values. This will be our $\log A$, and $\frac{\log 2}{\log A}$ will be the dimension of the \emph{Euclidean} statistically self-similar version of the Farey tree. 
\newline
\indent
Let us recall that we have $\frac{p_n}{q_n}(a_1\dots a_n)=\frac{1}{a_1+ \frac{1}{\dots \frac{1}{a_n}}}$ in the step $a_1+ \dots +a_n = N$ of the F-B partition. We estimate $q_n(a_1\dots a_n)$ according to Eq. \ref{eq:eq13} as $[c2^{\lambda _1}3^{\lambda _2} \dots ]^n$, where $\lambda_j$ is the proportion or frequency in which a coefficient $a_k$ equals $j$. Therefore $\log q_n = n[\log c + \lambda_1\log 2+\lambda_2\log 3+\dots] = n\log c +l_1\log 2+l_2\log 3 +...$, where $l_j$ is the total number of coefficients $a_k=j$. Then $\log q_n=n\log c+ \sum_j \log(a_j+1)$. We are in step $N=a_1+\dots+ a_n$. We also recall that we estimated length of segments as $\frac{1}{q_n^2}$, so, if we take the reciprocals and take logarithms we obtain $2\log q_n = 2 \{n\log c + \sum_j \log(a_j+1)\}; \sum_j a_j=N$. Before dividing by $N$, we will take averages of these values, in order to obtain $\log A, \frac{\log 2}{\log A}$ the dimension of the \emph{Euclidean} version of the Farey partition: we have to average the index "$n$" in a certain $N$-step; in order to average the values $\sum_j \log (a_j+1)$ we have to count first how many coefficients $a_j = 1 $ we have in step $N$, how many $a_j=2$\dots until $a_j=N$, which happens only once in that step. Then we can take averages and calculate $\log A$. The whole counting-and-averaging process, long and tedious, is done in the Appendix to Section 7. The result is $\log A = \log c + \frac{\log 2}{2^1} + \dots + \frac{\log (j+1)}{2^j}+ \dots$ .
\section{ ${\bf \log 2/\log A}$ is the Information Dimension for the $F-B$ Measure}
We want to compare this "Euclidean" dimensional version $\log 2/\log A$ of the $F-B$ measure with the entropy or information dimension for the $F-B$ measure in Sec. 5.5: $f(\alpha) = (-\frac{1}{2})\frac{\lambda_1\log\lambda_1+\dots}{\log c+\lambda_1\log 2+\lambda_2\log 3+\lambda_3\log 4+\dots},$ for $\lambda_j=\frac{1}{2^j}$. The denominators coincide. We have to compare $(-\frac{1}{2})(\lambda_1\log \lambda_1+\dots);\lambda_j=\frac{1}{2^j},$ with $\log 2$. We have $\log \lambda_j = \log \frac{1}{2^j} = -j\log 2$. So $-\frac{1}{2}(\lambda_1\log\lambda_1+\dots) = \frac{1}{2}(\frac{1}{2}\log 2+\frac{1}{2^2}2\log 2+\frac{1}{2^3}3\log 2+\dots) = \frac{\log 2}{2}\{\frac{1}{2}+\frac{2}{2^2}+\frac{3}{2^3}+\dots\}$. Let us calculate the expression within brackets. The Taylor expansion of $\frac{1}{(1-x)^2}$ is $1+2x+3x^2+\dots+(k+1)x^k+\dots$. For $x=\frac{1}{2}$ we have then $1+\frac{2}{2}+\frac{3}{2^2}+\dots+\frac{(k+1)}{2^k}+\dots=\frac{1}{(1-\frac{1}{2})^2} = 4$ which implies, dividing by $2$, $\frac{1}{2}+\frac{2}{2^2}+\frac{3}{2^3}+ \dots+\frac{k}{2^k}+\dots = 2$; so $2$ is the value of the expression within brackets, and $\log 2$ is the value of the numerator of $f(\alpha)$ above, which means that the expressions for $f(\alpha)$ and $\frac{\log 2}{\log A}$ above coincide, so $0.87038\dots$ is both the information or entropic dimension of the $F-B$ measure \emph{and} the dimension of the Euclidean version of the Farey tree partition. 
\section{Geometrical considerations, Conclusions and Conjectures }
\subsection{Geometrical considerations}
\subsubsection{Introduction}
For the content of this section we refer the reader to \emph{The Geometry of Farey Staircases} [Piacquadio, 2004] and to the corresponding references quoted there.
\newline\indent
There is a one-to-one connection between $F-B$ in $\mathds{R}$ and a certain non Euclidean geometry. Though we work on $I$, the $F-B$ interpolation is valid in any interval $[n,n+1], n \in \mathds{Z}$.
\newline\indent
Let $\mathds{H} = \{ z=x+\mathbf{\emph{i}}\emph{y} / (x,y) \in \mathds{R}^2, y >0\}$ be the upper half plane. We draw in $\mathds{H}$ the upper \emph{half} circles (centre in $\mathds{R}$) with endpoints in a pair of adjacent rationals $\frac{a}{b},\frac{a'}{b'}$ in any Nth $F-B$ partition. That is, we trace upper half circle (centre $\frac{1}{2})$ joining $0$ and $1$, then arc joining $0$ with $\frac{1}{2}$, then $\frac{1}{2}$ with $1$, \dots etc. in the Nth partition we trace $2^N$ small arcs joining adjacent rationals as endpoints. These arcs are geodesics in $\mathds{H}$. The three geodesics joining $\frac{a}{b}$ with $\frac{a'}{b'}$ (adjacent in an Nth $F-B$), $\frac{a}{b}$ with $\frac{a+a'}{b+b'}$, and $\frac{a+a'}{b+b'}$ with $\frac{a'}{b'}$ (in (N+1)th $F-B$), form a triangle in $\mathds{H}$. We have an infinite number of such triangles and, in the so-called Hyperbolic area measure, they all have the same \emph{area}.
\newline\indent
A rigid hyperbolic movement in $\mathds{H}$ is, by definition, a transformation $z \rightarrow \frac{a'z+a}{b'z+b}, z \in \mathds{H}, a, a', b, b'$ in $\mathds{Z}, $ det$\left( \begin{array}{ccc}
a' & a \\
b' & b 
 \end{array} \right)$ = 1. The set $\emph{U}$ of these movements can be seen as the multiplicative group of 2x2 matrices with integer entries and unit determinant. The \emph{triangles} above, do not only have the same hyperbolic \emph{area}, but are transformed into each other by rigid hyperbolic movements: by elements in $\emph{U}$: they are \textemdash hyperbolically\textemdash\space the \emph{same} triangle, moved here and there, to and fro. We do likewise in any interval $[n,n+1], n \in \mathds{Z}$.
\newline\indent
To the arcs described above, let us add \emph{vertical} lines $(n,\infty)$ with endpoint $n \in \mathds{Z}$ \textemdash which are also geodesics in $\mathds{H}$, the centre of the circle at infinity of $\mathds{R}$. On top of unit arc joining $0$ and $1$ \textemdash we will call it unit arc hereinafter\textemdash\space we have now another \emph{triangle}, the sides being vertical line $(0,\infty)$, unit arc, and vertical line $(1,\infty)$, vertices being $\infty$, $0$ and $1$. The same happens on top of arcs joining $n$ \& $n+1, n$ in $\mathds{Z}$. These \emph{new triangles} have the same \emph{area} as those above, and are interchangeable with them by elements in $\emph{U}$. All these \emph{non-overlapping} triangles \textemdash with finite or infinite vertices\textemdash\space cover $\mathds{H}$: they are a \emph{regular} tiling of $\mathds{H}$, and we will call it $\mathds{T}$ (\emph{T} for triangle and \emph{T}  for tiling).
\newline
\indent
In Sec. 2 we saw that, if $i = [a_1,\dots,a_n,\dots]$ and $[a_1\dots a_n]:=\frac{p_n}{q_n}$, then length of segment $[\frac{p_n}{q_n},\frac{p_{n+1}}{q_{n+1}}]$ is $\frac{1}{q_n q_{n+1}}$, which implies det$\left( \begin{array}{ccc}
p_{n+1} & p_n \\
q_{n+1} & q_n 
 \end{array} \right) = 1$ and $\left( \begin{array}{ccc}
p_{n+1} & p_n \\
q_{n+1} & q_n 
 \end{array} \right) \in \emph{U}$. This means that $\frac{p_n}{q_n}$ and $\frac{p_{n+1}}{q_{n+1}}$ are adjacent fractions in some $F-B$ partition, for all adjacent rationals $\frac{a}{b}, \frac{a'}{b'}$ in all $F-B$ partitions, $\frac{a}{b} < \frac{a'}{b'}$, have  $\left| \begin{array}{ccc}
a' & a \\
b' & b 
 \end{array} \right| = 1$, i.e.  $\left( \begin{array}{ccc}
a' & a \\
b' & b 
 \end{array} \right) \in \emph{U}$: there is a common structure in charge of $F-B$, continued fractions, and rigid movements in Hyperbolic Geometry; the algebraic group $\emph{U}$ being the common underlying principle.
\newline\indent
If  $\left( \begin{array}{ccc}
a' & a \\
b' & b 
 \end{array} \right) \in \emph{U}$, $0 < \frac{a}{b} < \frac{a'}{b'} < 1$, then  $\left( \begin{array}{ccc}
a' & a \\
b' & b 
 \end{array} \right)$ applied to unit segment shrinks $I$ into $[\frac{a}{b}, \frac{a'+a}{b'+b}]$, yielding the $F-B$ interpolation between adjacent $\frac{a}{b}$ and $\frac{a'}{b'}$; the length of the shrunk interval is $\frac{1}{b(b'+b)}$. Second row entries $b$ and $b'$ are non-zero and positive. Ditto when working in $[n,n+1]$ instead of $\emph{I}$, $n \in \mathds{N}$. When $n$ is negative, such entries are non-zero and negative. But other elements in $\emph{U}$ can have $b$ and $b'$ of different signs or zero, e.g. $u^* =  \left( \begin{array}{ccc}
n+1 & -1 \\
1 & 0 
 \end{array} \right)$ and $u =  \left( \begin{array}{ccc}
1-n & n \\
-1 & 1 
 \end{array} \right), n \in \mathds{Z}$. In both cases the denominator $b(b'+b)$ above is zero: $0(1+0)$ and $1(-1+1)$, respectively. Element $u^*$ transforms unit arc into vertical line $(\infty,n)$, whereas $u$ transforms unit arc into vertical $(n,\infty)$ \textemdash same line with different orientation, so $u$ and $u^*$ mirror each other\textemdash\space and unit segment into horizontal $[n,\infty)$: so $b(b'+b)=0$ for $u$ and $u^*$ shows that length $\frac{1}{b(b'+b)}$ of those lines \textemdash $u$(unit arch) and $u^*$(unit arch)\textemdash\space is infinity. Every $u$ in $\emph{U}$ has a $u^*$ mirror, related to $u$ in a technical way beyond the scope of this paper. An analogous analysis can be done to the translations $z\rightarrow z+n$, i.e.  $\left( \begin{array}{ccc}
1 & n \\
0 & 1 
 \end{array} \right) \in \emph{U}, n \in \mathds{Z}$. So, the correspondence between $F-B$ and $\emph{U}$ goes beyond the corresponding to non-zero-equal-sign-denominators of fractions in $I$, but extends to semicircular arcs with $F-B$ adjacent endpoints throughout $\mathds{R}$, and to vertical lines with endpoints in $\mathds{Z}$, i.e. to all geodesics delimiting all triangles in $\mathds{T}$. Notice that every such geodesic is obtained by applying each element of $\emph{U}$ to unit arc \textemdash which is the rationale for restricting the work to unit interval in the next sections. The $u/u^*$ mirror \emph{ambiguity} is avoided by joining smaller with larger values: 0 to 1 in unit arc or segment, $n$ to $\infty$ in the infinite lines. Other regular $F-B$ tesselations of $\mathds{H}$ aim to take care of this apparent ambiguity, but we stick to $\mathds{T}$, simpler to work with, and which embodies all geometric and metric properties of $\mathds{H}$, as well as defining, via the endpoints, the $F-B$ partition on $\mathds{R}$; which is the reason we have used the terms \emph{$F-B$ measure} and \emph{hyperbolic measure} in $\mathds{R}$ as interchangeable in the literature. 
\subsubsection{Equal $F-B$ measure of all intervals in Nth $F-B$ partition}
Let $i=[a_1,a_2,\dots] \in I$. Two matrices $L$ and $R$ \textemdash for \emph{left} and \emph{right}\textemdash\space in $\emph{U}$ can be constructed such that, applying $L$ $a_1$ times to $I$ means to $F-B$ interpolate $I$ $a_1$ times, each time choosing the \emph{left} interval in order to $F-B$ interpolate further. Ditto for $R$ and $a_2$\dots and so on. So $[a_1,a_2,\dots] = lim_{n\rightarrow \infty}L^{a_1}R^{a_2}\dots L^{2n+1}I$. Now $L$ and $R$ \textemdash left and right\textemdash are like, say, vectors $(0,1)$ and $(1,0)$ in $\mathds{R}^2$\textemdash horizontal and vertical: they carry the same weight, have the same "right to be present". So, e.g. $L^{a_1}R^{a_2}L^{a_3}$ will have the same $F-B$ or hyperbolic measure as $L^{a_1+ a_2+a_3}$ or $LR^{a_1+ a_2+a_3-1}$: all words written with $N$ letters $L$ and $R$ have the name weight or probability measure $\frac{1}{2^N}$ for each of the $2^N$ intervals in step $N$ of the $F-B$ partition.
\subsubsection{A deeper connection between $\mathds{H}$ and $F-B$}
Let us consider a vertical geodesic $G_i$ in $\mathds{H}$ with $i=[a_1\dots] \in I$  as endpoint. It cuts an infinity of triangles in $\mathds{T}$. Let us trace $G_i$ with a finger at its left side, from top to bottom. When crossing a triangle through a \emph{thin} part (only one vertex at left of $G_i$) we write \emph{T} for \emph{thin}, otherwise we write $\emph{F}$ (for \emph{fat}) \textemdash the tile at the very top of $G_i$ is \emph{T}, for technical reasons beyond this paper. We obtain an infinite word, letters $T$ and $F$: $T^{a_1}F^{a_2}\dots$ identical with $L^{a_1}R^{a_2}\dots$ in last section. This fact tightens the connection between continued fractions, $F-B$, and $\mathds{H}$. The main point here is that by naked eye observation, tracing $G_i$ with a finger, we can write directly any $i$ in its continued fraction, without any calculation. Let us recall (Sec. 5) that irrationals with the same $\alpha$-concentration are those with, roughly, the same average over the $a_j$ values: this can be verified by \emph{looking} at $G_i$'s: cardinality of tiles in $\mathds{T}$ with adjacent $T$'s or $F$'s should be \textemdash statistically\textemdash\space the same. Also, knowledge on the $a_j$'s of an irrational $i$, implies knowing the classification of said $i$ (Bruno, Jarnik, Liouville\dots) needed by physicists to study circle maps or optoelectronic phenomena [Piacquadio \& Rosen, 2007]. 
\newline\indent
Now: suppose that we have the ordinary half plane $\mathds{R}^2$ with an ordinary \emph{regular} tiling, all tiles interchangeable by rigid Euclidean motions. Notice that no geodesics \textemdash vertical or otherwise\textemdash\space crossing the tiles with endpoint in an irrational $i \in \mathds{R}$ will yield those tools to classify said $i$ according to the criteria needed by physicists, whereas $\emph{any}$ geodesic in $\mathds{H}$ with endpoint in $i$ \textemdash not only the vertical $G_i$\textemdash\space will yield such classification.
\subsubsection{Fundamental differences between Euclidean $\mathds{R}^2$ and $\mathds{H}$} 
So far, we stressed the tight connection between $F-B$, continued fractions and $\mathds{H}$ (cum $\emph{U}$ cum $\mathds{T}$), with an emphasis in $\mathds{H}$. And, at the end of last section, we pointed out like a divorce between upper half planes $\mathds{R}^2$ and $\mathds{H}$. Such differences run deep indeed: we can have $\mathds{R}^2$ regularly tesselated by triangles, squares, hexagons\dots period, whereas it is a most satisfying exercise to transform $\mathds{H}$ into the Poincare circle $\mathds{P}$, to choose, say, five or eight consecutive geodesics, and tesselate $\mathds{P}$ (hence $\mathds{H}$) with regular pentagons, octagons,\dots etc \textemdash an impossible endeavor in $\mathds{R}^2$. Opposite characteristics are easy to observe even at the level of $\mathds{R}$: when we write $x=0, a_1\dots a_n 00 \dots \in I$, we know that $x$ is rational, whereas in $F-B$ hyperbolic style a rational in $I$ is written $[a_1,\dots a_n,\infty,\infty,\dots]$. Likewise $x=0,a_1 a_2 a bab \dots$ is rational, whereas $[a_1,a_2,a,b,a,b,\dots]$ belongs to the most irrational category\dots
\subsubsection{Analogies between the two measures, Euclidean and Hyperbolic}
The list above of apparently irreconciliable differences between the two measures is by no means complete, for many more are pointed out in the literature. Some analogies, instead, have been noticed in [Piacquadio and Cesaratto, 2001], and they begin to appear, obscurely, through multifractal analysis. 
\newline\indent
In Sec. 7 we take the Euclidean length of the $2^N$ $F-B$ intervals in step $N$ with a common Euclidean ruler. We obtain a list of $2^N$ values, of which we take logarithms. Some values are larger, some are smaller, so we take their average, which yields a single statistical contractive value $\frac{1}{A} \in (0,\frac{1}{2})$, so $\frac{\log 2}{\log A}$ is the statistical self similar dimension of the $F-B$ partition. In the ternary of Cantor $\frac{1}{A} = \frac{1}{3}$, and we have $2^N$ subfractals \textemdash in segments of Euclidean length $\frac{1}{3^N}$\textemdash\space interchangeable by rigid Euclidean movements, \dots all of which happens with the single contractor $\frac{1}{A}$ above: it yields a \textemdash Euclideanly\textemdash\space self-similar fractal, a statistical counterpart of the $F-B$ partition. But in Sec. 8 we learn that its dimension $\frac{\log 2}{\log A}$ is the same $d \sim 0.870\dots$: here is a deep contact between Euclidean and Hyperbolic geometries.
\subsection{Conclusions and Conjectures}
The two measures, Euclidean and Hyperbolic meet in a very specific dimension: $ 0.8700 \pm 0.0004$. This value of $f(\alpha)$, the entropic or information dimension, corresponds to the fractal where the Hyperbolic measure is concentrated \textemdash whereas the dimensional Euclidean perspective \textquotedblleft sees" the Farey Tree partition as having this specific dimension, instead of dimension 1. This universal number, therefore, is strongly perched on, and comfortably accommodated in, the intersection of the two measures. How does it appear in the dynamics of the Circle Map? For just a moment let us suppose we understand that the Circle Map acts as a black box: the input is the \textquotedblleft $y$"--vertical axis in the Devil's Staircase associated with the map: the entire unit segment is there, the input is, dimensionally, 1. The output is the selected subfractal $\Omega \subset I$ in the horizontal axis (associated with the circle map staircase) of dimension $0.870\dots$ . This black box seems to act as a dimensional spaghetti percolator: the output, what is retained, is, Hyperbolically, that set where the measure is dimensionally concentrated, yielding full information on such measure. This would be the thick fat spaghetti, whereas what is lost, what oozed through the percolator holes is the very small stuff: herbs, salt, fine flour, seasoning, small particles that came with the spaghetti in the input,\dots which do not yield much information, do not concentrate the measure.
\newline 
\indent
From the Euclidean point of view, the \emph{whole} of the input is dimensionally retained in the percolator, for the Euclideanly self-similar version of the Farey partition has exactly this $0.870\dots$ dimension. 
\newline
\indent
Let us assume we accept that the circle-map-Devil-Staircase black box acts as such a percolator: it retains the concentration of information. Then, the universal character of this numerical constant might be clear: changing the \textquotedblleft sine" function in the circle map by another reasonably smooth function that draws the circle, would mean changing a percolator by another of a slightly different form, say, an enamelled one with little circular holes, by a wire net one with adjacent square holes: the same spaghetti would remain trapped, the same output would be obtained, the same tiny particles lost.
\newline
\indent
Why and how the circle-map-Devil-staircase black box acts as such a measure percolator, however, still remains, for us, a mystery.

\section{References}
Bak, P. [1986] \textquotedblleft The devil's staircase", \emph{Phys. Today} December, 38-45.
\newline
Cawley, R. \& Mauldin, R.D. [1992] \textquotedblleft Multifractal decompositions of Moran fractals", \emph{Adv. Math.} \textbf{92}(2), 196-236.
\newline
Cesaratto, E. \& Piacquadio, M. [1998] \textquotedblleft Multifractal formalism of the Farey partition", \emph{Revista de la Uni$\acute{\mathrm{o}}$n Matem$\acute{\mathrm{a}}$tica Argentina} \textbf{41}(2), 51-66.
\newline
Falconer, K. [1990] \emph{Fractal Geometry Mathematical Foundations and Applications} (John Wiley, Chichester, New York), Chap. 17.
\newline
Good, I. J. [1941] \textquotedblleft The fractional dimensional theory of continued fractions", \emph{Proc. Camb. Phil. Soc. } \textbf{37}, 199-228.
\newline
Hensley, D. [1996] \textquotedblleft A polynomial time algorithm for the Hausdorff dimension of continued fraction Cantor sets", \emph{J. Numb. Th.} \textbf{58}, 9-45.
\newline
Jarnik, V. [1928-1929] \textquotedblleft Zur metrischen Theorie der diophantischen Approximationen", \emph{Prace Mat.-Fiz.}, 91-106.
\newline
Jarnik, V. [1929] \textquotedblleft Diophantischen Approximationen und Hausdorffsches Mass", \emph{Mat. Sbornik} \textbf{36}, 371-382.
\newline
Jensen, M. H., Bak, P. \& Bohr, T. [1984] \textquotedblleft Transition to chaos by interaction of resonances in dissipative systems. I. Circle maps", \emph{Phys. Rev. A} \textbf{30}(4), 1960-1969.
\newline
Piacquadio Losada, M. [2004] \textquotedblleft The geometry of Farey staircases", \emph{Int. J. Bifurcation and Chaos } \textbf{14}(12), 4075-4096.
\newline
Piacquadio, M. \& Cesaratto, E.  [2001] \textquotedblleft Multifractal spectrum and thermodynamical formalism of the Farey tree", \emph{Int. J. Bifurcation and Chaos} \textbf{11}(5), 1331-1358.
\newline
Piacquadio, M. \& Rosen, M.  [2007] \textquotedblleft Multifractal spectrum of an experimental (video feedback) Farey tree", to appear in \emph{J. Stat. Phys.} \textbf{127}(4), 783-804.
\newline
Riedi, R. \& Mandelbrot, B. [1997] \textquotedblleft The inversion formula for continuous multifractals", \emph{Adv. Appl. Math.} \textbf{19}, 332-354.
\newline
Riedi, R. \& Mandelbrot, B. [1998] \textquotedblleft Exceptions to the multifractal formalism for discontinuous measures", \emph{Math. Proc. Camb. Phil. Soc.} \textbf{123}, 133-157.
\newline
Series, C. \& Sinai, Y. [1990] \textquotedblleft Ising models on the Lobachevsky plane", \emph{Commun. Math. Phys.} \textbf{128}, 63-76.
\newline
Weisstein, E. W. [2005] \textquotedblleft Devil's staircase", \emph{MathWorld - A Wolfram Web Resource}, http://mathworld.wolfram.com/DevilsStaircase.html.
\section{Appendix to Section 3.2}
\label{sec:app_a}
We have to extremize $-\frac{1}{\log n_0}(\lambda_1\log\lambda_1+\dots)-\Lambda\frac{-1}{\log n_0}(\lambda_1\log p_1+\dots-\alpha)+\mu(\lambda_1+\dots-1)$, so the derivative with $\lambda_j$ as variable: $-\frac{1}{\log n_0}(1+\log\lambda_j) -\Lambda\frac{-1}{\log n_0}\log p_j+\mu=0$ implies $1+\log\lambda_j-\Lambda\log p_j+\mu^*=0$.
\newline
\indent
Subtracting the equation corresponding to $\lambda_1$ we have $\log\frac{\lambda_j}{\lambda_1}-\Lambda\log\frac{p_j}{p_1}=0,$ or $\frac{\lambda_j}{\lambda_1}= (\frac{p_j}{p_1})^\Lambda$, $j=2\rightarrow n_0$. Then $\lambda_j=\frac{\lambda_1}{p_1^\Lambda}p_j^\Lambda$, $j=1 \rightarrow n_0$, and from $\sum_j\lambda_j=1$ we obtain $\lambda_1=\frac{p_1^\Lambda}{\sum_j p_j^\Lambda}=\frac{p_1^\Lambda}{\mathcal{E}},$ hence 
\begin{equation}  \label{eq:eq3}
\lambda_j= \frac{\lambda_1}{p_1^\Lambda}p_j^\Lambda= \frac{p_1^\Lambda}{\mathcal{E}}\cdotp\frac{1}{p_1^\Lambda}p_j^\Lambda = \frac{p_j^\Lambda}{\mathcal{E}} \qquad \forall j
\end{equation}

The condition $-\frac{1}{\log n_0}\sum_j\lambda_j\log p_j=\alpha$ permits knowing the value of $\Lambda$:
\begin{equation} \label{eq:eq4}
\Lambda=\Lambda(\alpha), \textrm{ or } \alpha=\alpha(\Lambda)
\end{equation}
With 
\begin{equation} \label{eq:eq5}
f(\alpha)=\frac{-1}{\log n_0}\sum_j\lambda_j\log\lambda_j;\qquad 
\alpha = \frac{-1}{\log n_0}\sum_j\lambda_j\log p_j;\qquad
\lambda_j=\frac{p_j^\Lambda}{\mathcal{E}}
\end{equation}
we can calculate $f'(\alpha)= \frac{df}{d\alpha}= \frac{df/d\Lambda}{d\alpha/d\Lambda}= \frac{\sum(1+\log\lambda_j)\lambda'_j}{\sum \lambda'_j\log p_j} = \frac{\sum(\log\lambda_j)\lambda'_j}{\sum \log p_j\lambda'_j}$ since $\sum\lambda'_j=0$. Therefore $f'(\alpha)= \frac{\sum \lambda'_j\{\Lambda\log p_j- \log \mathcal{E}\}}{\sum\log p_j\lambda'_j}=\Lambda$, again since $\sum_j\lambda'_j=0$.  
\section{Appendix 1 to Section 4}
\label{sec:app_b}
We have to extremize the function $\frac{\lambda_1\log\lambda_1+\dots}{\lambda_1\log c_1+\dots} - \Lambda(\frac{-\log n_o}{\lambda_1\log c_1+\dots}-\alpha)+ \mu(\lambda_1+\dots - 1)$. The derivative of this function, $\lambda_j$ as variable, equated to zero yields: $ \frac{(1+\log\lambda_j)(\lambda_1\log c_1+\dots)-(\lambda_1\log\lambda_1+\dots)\log c_j}{(\lambda_1\log c_1+\dots)^2} + \Lambda\frac{(-\log n_o)}{(\lambda_1\log c_1 + \dots)^2}\log c_j+\mu=0$, hence $\log\lambda_j(\lambda_1\log c_1+\dots)- (\lambda_1\log\lambda_1+\dots)\log c_j-\Lambda\log n_0\log c_j+ \mu(\lambda_1\log c_1+\dots)^2+(\lambda_1\log c_1+\dots)=0$ or $\log\lambda_j(\lambda_1\log c_1+\dots)- \log c_j(\lambda_1\log\lambda_1+\dots+\Lambda\log n_0) = $ a constant independent of $j$. Subtracting the corresponding equality for $j=1$, writing $\lambda_1\log c_1+\dots=( ),$ and $\lambda_1\log\lambda_1+\dots+\Lambda\log n_0=[ ]$ for short, we have $( )\log\frac{\lambda_j}{\lambda_1}-\log\frac{c_j}{c_1}[ ]=0,$ or $\frac{\lambda_j}{\lambda_1}=(\frac{c_j}{c_1})^{[ ]/( )};$ $\lambda_j= \frac{\lambda_1}{c_1^{[]/()}}c_j^{[]/()}$ with $j=1 \rightarrow n_0.$

As in Sec. 3  we use $\sum_j \lambda_j = 1$, we obtain $\lambda_1$ and then $\lambda_j$, with the result 
\begin{equation} \label{eq:eq6}
	\lambda_j = \frac{c_j^{[]/()}}{\mathcal{E}} := \frac{c_j^\Xi}{\mathcal{E}}
\end{equation}
which bears a resemblance to Eq. \ref{eq:eq5}.
\newline
\indent
Next, we want to calculate $f'(\alpha) = \frac{df}{d\alpha}$. We write $f(\alpha) = \frac{\lambda_1\log\lambda_1+\dots}{\lambda_1\log c_1+\dots} := \frac{\mathrm{num}}{\mathrm{den}}$, for short. Then $\alpha=\frac{-\log n_0}{\mathrm{den}}.$ Therefore $f('\alpha)=\frac{df}{d\alpha}= \frac{df/d\Xi}{d\alpha/d\Xi}= \frac{\mathrm{ \{num' den - num \cdot den'\}/den}^2}{-\log n_0 (-1/\mathrm{den}^2)\mathrm{den'}} = \frac{1}{\log n_0}\frac{[(1+\log\lambda_1)\lambda_1'+\dots](\lambda_1\log c_1\dots)- (\lambda_1\log\lambda_1+\dots)(\lambda_1'\log c_1+\dots)}{\lambda_1'\log c_1+\dots}$.
Since $\sum\lambda_j'=0$ we have $f'(\alpha)=\frac{1}{\log n_0}\{\frac{\lambda_1'\log\lambda_1+\dots}{\lambda_1'\log c_1+\dots}(\lambda_1\log c_1+\dots)- (\lambda_1\log\lambda_1+\dots)\}$. Now, $\log\lambda_j = \Xi\log c_j-\log \mathcal{E}$ implies $\lambda_1'\log\lambda_1+\dots = \sum_j\Xi\log c_j\lambda_j' - \log \mathcal{E}\sum_j\lambda_j'=\Xi(\lambda_1'\log c_1+\dots).$ Hence,
\begin{eqnarray} \label{eq:eq7}
\nonumber
f'(\alpha)= \frac{1}{\log n_0}\{\Xi(\lambda_1\log c_1+\dots)- (\lambda_1\log\lambda_1+\dots)\} = \\
\frac{1}{\log n_0}\{\Xi(\lambda_1\log c_1+\dots) - (\lambda_1\Xi\log c_1+\dots)+ \log \mathcal{E} (\lambda_1+\dots)\}= 
\frac{\log \mathcal{E}}{\log n_0}
\end{eqnarray}
\section{Appendix 2 to Section 4}
\label{sec:app_c}
 Variable $\bar{\alpha}$, the natural independent variable of $\bar{f}$ is 
\begin{equation} \label{eq:eq8}
\bar{\alpha}=\frac{1}{\alpha},
\end{equation}
$\alpha$ the \textquotedblleft old" concentration from Sec 3. The inversion formula of Riedi and Mandelbrot [1997 and 1998] says that the new inverted spectrum $\bar{f}$ is related to the old $f: \bar{f}(\alpha)=\alpha f(\frac{1}{\alpha})$. From this we have $\bar{f'}(\alpha) = f(\frac{1}{\alpha}) + \alpha f'(\frac{1}{\alpha})\frac{-1}{\alpha^2} = f(\frac{1}{\alpha}) - \frac{1}{\alpha}f'(\frac{1}{\alpha})$, and from above this becomes $\bar{q} = \bar{f'}(\bar{\alpha})= \bar{f'}(\frac{1}{\alpha}) = f(\alpha)-\alpha f'(\alpha) = f(\alpha) - \alpha q = -(q\alpha - f(\alpha)) = -\tau(q)$, that is
\begin{equation} \label{eq:eq9}
-\tau(q)=\bar{q} \qquad \forall q
\end{equation} 
where $\tau$ and $q$ are \textquotedblleft old" parameters. Applying again the same criterion we have $\overline{(\bar{q})} = - \bar{\tau}(\bar{q}) = - \bar{\tau}(-\tau(q))$, that is $\overline{\bar{q}} = -\bar{\tau}(-\tau(q)).$ But $\overline{\bar{f}} = f$ implies $\overline{\bar{q}} = q$, so we have
\begin{equation} \label{eq:eq10}
-\bar{\tau}(-\tau(q))= q \qquad \forall q \qquad \mathrm{or} -\tau(-\bar{\tau}(\bar{q})) = \bar{q}  \qquad \forall \bar{q} 
\end{equation}
Let us go back to Eq. \ref{eq:eq7}: in our new notation:
\begin{eqnarray} \label{eq:eq11}
\nonumber
 \frac{d\bar{f}}{d\bar{\alpha}} = \frac{\log \mathcal{E}}{\log n_0} =
 \frac{-N\log(c_1^\Xi+ \dots)}{N\log\frac{1}{n_0}} = \frac{-\log(c_1^\Xi+\dots)^N}{\log\frac{1}{n_0^N}} = \\ \frac{-\log\sum_{r_1+\dots=N}c_1^{\Xi r_1}\dots c_{n_0}^{\Xi r_{n_0}}}{\log(\frac{1}{n_0})^N} = - \frac{\log\sum_{r_1+\dots = N}(c_1^{r_1}\dots c_{n_0}^{r_{n_0}})^\Xi}{\log (\mathrm{length})} = - \tau(\Xi) 
\end{eqnarray}
for we have the case \textquotedblleft equal lengths and different probabilities" given by contractors $c_j,\dots$ so we are in Sec. 3 with the very definition of $\tau$.
\newline
\indent
Now, let us focus on variable $\Xi = \frac{[]}{()} = \frac{\lambda_1\log\lambda_1+ \dots+\Lambda
\log n_0}{\lambda_1\log c_1 + \dots } = \frac{\lambda_1\log\lambda_1+\dots}{\lambda_1\log c_1+\dots}  -\Lambda \frac{-\log n_0}{\lambda_1\log c_1+\dots} = \bar{f}(\bar{\alpha}) - \Lambda\bar{\alpha} = -(\Lambda\bar{\alpha} - \bar{f}(\bar{\alpha})) = -\bar{\tau}(\Lambda)$, so Eq. \ref{eq:eq11} now reads $-\tau(\Xi) = -\tau(-\bar{\tau}(\Lambda)) = \Lambda$   $\forall\Lambda$ from Eq. \ref{eq:eq10}. Hence, we have our derivative $\bar{f'}(\bar{\alpha}) = \Lambda$,  as was the case of \textquotedblleft equal lengths" in Sec. 3. It means that $\bar{q}=\Lambda$ in Sec. 4, as $q=\Lambda$ in Sec. 3, in both cases $\Lambda$ being the Lagrange indeterminate coefficient linking $f(\alpha)$ with $\alpha$.
\section{Appendix to Section 5.2}
\label{sec:app_d}
We need, now, to estimate constant $c \in (0,1)$. We will adapt a reasoning that we used elsewhere [Piacquadio, 2004] in order to apply the methods in Sec. 4. 
\newline
\indent
Let $E_k=\{ i = [a_1\dots a_j\dots]/a_j \leq k$ $\forall j \}$, $k \in \mathds{N}$. We will estimate the Hausdorff dimension $d_H(E_k)$ by considering finite sequences $[a_1\dots a_n]$, $a_j\leq k$ (later $n$ will tend to infinity), and considering  \textemdash as above\textemdash\space the $\lambda_j$ as the frequency in which the $a$'s are equal to $j$. For a certain choice of $\lambda_1\dots\lambda_k$ we have the corresponding dimension given by $\frac{\log(1/\lambda_1^{\lambda_1}\dots \lambda_k^{\lambda_k})^n}{\log (c 2^{\lambda_1}\dots (k+1)^{\lambda_k})^{2n}}$, simply repeating the processes above. So, $d_H(E_k)$ will be obtained by finding the key set of frequencies $\lambda_j$ for which 
\begin{equation} \label{eq:eq14}
-\frac{1}{2}\frac{\lambda_1\log\lambda_1+\dots}{\log c+\lambda_1\log 2+\dots+\lambda_k\log(k+1)} := -\frac{1}{2}\frac{\mathrm{num}}{\mathrm{den}}
\end{equation}
reaches its maximum.
\newline
\indent
Again $\frac{d}{d\lambda_j}(\frac{\mathrm{num}}{\mathrm{den}})=0$ implies $(1+\log\lambda_j)\mathrm{den}=\mathrm{num}\log(j+1)$, that is $1+\log\lambda_j= \frac{\mathrm{num}}{\mathrm{den}}\log(j+1)$, or, from Eq. \ref{eq:eq14} 
\begin{equation} \label{eq:eq15}
1+\log\lambda_j= -2 d_H(E_k)\log(j+1) := a\log(j+1).
\end{equation}
Following already well trodden steps, we obtain $\log\frac{\lambda_j}{\lambda_1} = \log [\frac{(j+1)}{2}]^a$ from above, from which $\lambda_j = \frac{\lambda_1}{2^a}(j+1)^a$. With $\sum_j \lambda_j=1$ we obtain 
\begin{equation} \label{eq:eq16}
\lambda_1 = \frac{2^a}{\mathcal{E}}, \qquad \mathrm{and} \qquad \lambda_j = \frac{(j+1)^a}{\mathcal{E}}.
\end{equation}
Now let us examine the value of 
\begin{equation} \label{eq:eq17}
a = -2 d_H(E_k).
\end{equation}
As $k \rightarrow \infty$, $d_H(E_k)$ must tend to unity, as $E_k$ tends to encompass every $i=[a_1\dots a_j\dots]$ regardless of the size of the $a$'s. In fact, Jarnik [1928; 1929] proved $1-d_H(E_k)=O(\frac{1}{k})$. Therefore, as $k$ grows, $a \cong -2$. Hence, $\mathcal{E} \cong \sum_1^\infty(j+1)^{-2}=\frac{\pi^2}{6}-1 := c_\pi$. Thus Eq. \ref{eq:eq16} becomes  $\lambda_j = \frac{(j+1)^-2}{c_\pi}$, $k$ large, and Eq. \ref{eq:eq14} becomes (\emph{always} $k$ large) $- 2 d_H(E_k) \cong -2 = \frac{\mathrm{num}}{\mathrm{den}} = \frac{\sum_j\frac{(j+1)^{-2}}{c_\pi}\{-2\log(j+1)-\log c_\pi \} }{\log c +\frac{1}{c_\pi}\sum_j(j+1)^{-2}\log(j+1)} := \frac{-2 \sum \sum \sum - \frac{1}{c_\pi}\sum(j+1)^{-2}\log c_\pi}{\log c + \sum \sum \sum}$ from which $ -2 \log c = -\frac{1}{c_\pi}\sum(j+1)^{-2}\log c_\pi = -\log c_\pi$, that is $c=\sqrt{c_\pi}$. 
\section{Appendix to Section 5.4}
\label{sec:app_e}
With the same notation as in Sec 5.3, we want to extremize $\frac{\log(\frac{1}{\lambda_1^{\lambda_1}\dots \lambda_l^{\lambda_k}})^n}{ \log \{\{([c 2^{\lambda_1}\dots (k+1)^{\lambda_k}]^n)^2\}\}} - \Lambda(\frac{\log \frac{1}{2^{\sum a_j}}}{\log(1/\{\{\}\})} - \alpha )$, with condition $\sum \lambda_j = 1$. Proceeding as in the Euclidean case, we have to find extremes of $-\frac{1}{2}\frac{\lambda_1\log\lambda_1+\dots}{\log(c 2^{\lambda_1}\dots)}- \Lambda\frac{-\log 2 \sum_j j\lambda_j}{-2\log( c 2^{\lambda_1}\dots)}$, that is, the extremes of $\frac{\lambda_1\log\lambda_1+\dots+\Lambda\log 2\sum_j j\lambda_j}{\log c+\lambda_1\log 2+\dots} := \frac{\mathrm{num}}{\mathrm{den}}$. We equate the derivative of this function (variable $\lambda_j$) to zero, and the difficulties in the (apparent) differences with the Euclidean case begin to appear: $((\log\lambda_j+1)+j\Lambda\log 2)\mathrm{den} = \mathrm{num}\log (j+1)$, so $1+\log\lambda_j+j\Lambda\log 2 = \frac{\mathrm{num}}{\mathrm{den}}\log(j+1)$, with $\frac{\mathrm{num}}{\mathrm{den}} = -2f(\alpha)+ \Lambda\frac{\log 2\sum_j j\lambda_j}{\log c+\lambda_1\log 2+\dots} = -2f(\alpha)+2\Lambda\alpha = -2(f(\alpha)-\Lambda\alpha) = 2\tau(\Lambda)$. Therefore, proceeding as before, we have $\log\frac{\lambda_j}{\lambda_1}+ (j-1)\Lambda\log 2 = 2\tau(\Lambda)\log\frac{(j+1)}{2}$, or $\log\frac{\lambda_j}{\lambda_1} = \log(\frac{j+1}{2})^{2\tau(\Lambda)} - \log 2^{\Lambda(j-1)}$ which means $\lambda_j = \lambda_1(\frac{j+1}{2})^{2\tau}\cdotp \frac{1}{2^{\Lambda(j-1)}}$, and the equality is valid for $j=1$ as well. With $\sum_j\lambda_j =1$ we obtain, as before, the value of $\lambda_1$ and then that of $\lambda_j$:
\begin{equation} \label{eq:eq22}
\lambda_j = \frac{(j+1)^{2\tau(\Lambda)}/2^{\Lambda(j-1)}}{\mathcal{E}}.
\end{equation}
\section{Appendix to Section 7}
\label{sec:app_f}
In order to calculate $\log A$ we need to closely study the nature of the $n$'s and $a_j$'s in a certain $N$ step.
\subsection{Rewriting the tree}
We start with the first interpolations of the tree: 
\newline
\rowheight=0.5pc
\xytree{
& & & \xynode[0]{$[\frac{0}{1},\frac{1}{1}]$} \\ 
& & & \xynode[-2,2]{$\frac{1}{2}$} \\
& \xynode[-1,1]{$\frac{1}{3}$}& & & & \xynode[-1,1]{$\frac{2}{3}$} \\
\xynode{$\frac{1}{4}$} &  & \xynode{$\frac{2}{5}$} & & \xynode{$\frac{3}{5}$} & &  \xynode{$\frac{3}{4}$}
}
\newline
\indent
Let us express, in terms of continued fractions, the values $\frac{2}{3},\frac{2}{5},\frac{3}{5}$ and $\frac{3}{4}$: $\frac{2}{3} = \frac{1}{\frac{3}{2}} = \frac{1}{1+\frac{1}{2}}$, so the $a_k$ involved are 1 and 2;
$\frac{2}{5} = \frac{1}{\frac{5}{2}} = \frac{1}{2 + \frac{1}{2}}$, so the $a_k$ involved are 2 and 2; 
 $\frac{3}{5} = \frac{1}{\frac{5}{3}} = \frac{1}{1+\frac{2}{3}} = \frac{1}{1+\frac{1}{\frac{3}{2}}} = \frac{1}{1+\frac{1}{1+\frac{1}{2}}}$, so $a_k$ are 1, 1, and 2. And $\frac{3}{4} = \frac{1}{\frac{4}{3}} = \frac{1}{1+\frac{1}{3}}$, with $a_k$ being 1 and 3. So the tree above can be rewritten thus:
\xytree{
& & & \xynode[-2,2]{$[2]$} \\
& \xynode[-1,1]{$[3]$}& & & & \xynode[-1,1]{$[1,2]$} \\
\xynode{$[4]$} &  & \xynode{$[2,2]$} & & \xynode{$[1,1,2]$} & &  \xynode{$[1,3]$}
}\newline
So the step can be seen as the sum of the $a_k$ involved inside brackets in each horizontal line:
\newline
\xytree{
\xynode{N=2}  & \xynode{\dots} & \xynode{\dots} & \xynode{\dots} & \xynode[-2,2]{[2]} \\
\xynode{N=3}  & \xynode{\dots} & \xynode[-1,1]{[3]} & & \xynode{;} & & \xynode[-1,1]{[1,2]} & \xynode{\dots} & \xynode{\dots} & \xynode{1+2=3} \\
\xynode{N=4}  & \xynode{[4]} & \xynode{;} & \xynode{[2,2]} & \xynode{;} & \xynode{[1,1,2]} & \xynode{;} & \xynode{[1,3]} & \xynode{\dots} & & \xynode{2+2=1+1+2=1+3=4} \\
}
\newline \newline
and so on. Let us examine the minitree 
\newline
\xytree{
& \xynode[-1,1]{[1,2]}\\
\xynode{[1,1,2]} && \xynode{[1,3]} \\
}
\newline
\newline
We observe that the cypher 1 in $[1,2]$ appears in both the daughter branches, whereas the \emph{last} $a_k$ in $[1,2]$, i.e. 2, unfolded thus: $[\dots 2-1,2]$ or $[\dots 1,2]$.
\newline
\indent
This is general, as we see by examining the other minitrees, or by extending the tree to $N=5,6,\dots$. The process: $[a_1, \dots , a_{n-2}, a_{n-1}]$ in step $N-1 = a_1+\dots+a_{n-1}$ generates $[a_1,\dots a_{n-2},a_{n-1}+1]$ and $[a_1,\dots a_{n-2},a_{n-1}-1,2]$ in step $N$. 
\newline
\indent
We need to estimate the average of all $\frac{\log q_n}{N}$ in a certain step $N$. The tree above has \emph{only} the \emph{new} elements which appear in step $N$; we will call this the restricted tree \textemdash restricted only to the \emph{new} elements in the step. The tree with \emph{all} elements will be called the complete tree. The elements in the $N^{th}$ horizontal line of the restricted tree will be the restricted elements in step $N$.
\subsection{Averaging index \textquotedblleft $n$" in an $N$-step}
Since $\log q_n$ is estimated by $n\log c + \sum_j \log(a_j+1)$ we will start by averaging all $\frac{n\log c}{N}$ involved in step $N$. We start by adding up all the \textquotedblleft $n$'s" in a restricted $N$-step.
Let us enlarge the restricted tree a bit more: 
\newline
\begin{tabular}{l@{}r@{}l@{}l}
$N=2$ &  & $\longrightarrow$ & $[2]$ \\
$N=3$ &  & $\longrightarrow$ & $[3];[1,2]$ \\
$N=4$ &  & $\longrightarrow$ & $[4];[2,2];[1,3];[1,1,2]$ \\
$N=5$ &  & $\longrightarrow$ & $[5];[3,2];[2,3];[2,1,2];[1,4];[1,2,2];[1,1,3];[1,1,1,2]$ \\
\dots\\
\end{tabular}
\newline
\newline
The corresponding values $n(N)$, for step $N$, i.e. the lengths of integers $a_k$ inside brackets are:
\newline
\newline
\begin{tabular}{l c l}
$N$ & & $n(N)$ \\
2 & \dots & 1 \\
3 & \dots & 1,2 \\
4 & \dots & 1,2,2,3 \\
5 & \dots & 1,2,2,3,2,3,3,4 \\
\dots && \\
\end{tabular}
\newline
\newline
Rearranged, these numbers are:
\newline
\newline
\begin{tabular}{l c l}
$N$ & & $n(N)$ \\
2 & \dots & 1 \\
3 & \dots & 1,2 \\
4 & \dots & 1,2,2,3 \\
5 & \dots & 1,2,2,2,3,3,3,4 \\
6 & \dots & 1,2,2,2,2,3,3,3,3,3,3,4,4,4,4,5 \\
\dots & & \\
\end{tabular}
\newline
\newline
By simple observation we see that the lengths $n(N)$ vary from 1 to $N-1$, and that their repetition follows the combinatorial numbers in the Pascal triangle of order $N-2$. So the sum of all the $n(N)$ in the restricted $N$ step is ${N-2\choose 0} \cdot 1 + {N -2 \choose 1}\cdot 2 + \dots + {N-2 \choose N-2}(N-1) = \sum_{j=0}^{N-2} {N-2 \choose j}(j+1)$, and we claim that this sum is $N 2^{N-3}$. It is obviously true for $N=2$, the first $N$ of our account. Let us assume the validity of 
\begin{equation} \label{eq:eq26}
\sum_{j=0}^{N-2} {N-2 \choose j}(j+1) = N 2^{N-3}
\end{equation}
for a certain value of $N$, and let us infer the validity of said equation for $N+1$. For short, we write $N-2=k$, so Eq. \ref{eq:eq26} becomes $\sum_{j=0}^k {k \choose j}(j+1) = (k+2)2^{k-1}$. We have
\begin{eqnarray}
\nonumber
\sum_{j=0}^{k+1} {k +1 \choose j}(j+1) = 
1 + \sum_{j=1}^k {k + 1\choose j}(j+1) + (k+2) = \\
\nonumber
1 + \sum_{j=1}^k [{k \choose j} + {k \choose j-1} ](j+1) + (k+2) =\\ 
\nonumber
1 + \sum_{j=1}^k {k \choose j}(j+1) + \sum_{j=1}^k {k \choose j-1}j +\sum_{j=1}^k {k \choose j-1} + (k+2) =\\ 
\nonumber
\sum_{j=0}^k {k \choose j}(j+1) + \sum_{j-1=0}^{j-1=k-1} {k \choose j-1}[(j-1)+1]+(k+1)+ \sum_{j=0}^k {k \choose j}=\\
\nonumber
2\sum_{j=0}^k {k \choose j}(j+1) + 2^k = 2(k+2)2^{k-1}+2^k=(k+3)2^k
\end{eqnarray}
which is Eq. \ref{eq:eq26} with k replaced by $k+1$. Therefore, the sum of all $n(N)$ in the restricted step $N$ is $N 2^{N-3}$.
\newline
\indent
Next, we need the sum of \emph{all} $n(N)$ in step $N$: 
\begin{eqnarray}
\nonumber
2\cdot 2^{2-3} + 3\cdot 2^{3-3} + \dots + N2^{N-3} = \frac{1}{4} \{2x^{2-1}+3\cdot x^{3-1}+\dots+Nx^{N-1}\} \arrowvert_{x=2} \\
\nonumber
= \frac{1}{4}(x^2+x^3+\dots+x^N)' \arrowvert_{x=2} = \frac{1}{4}\{x^2(1 + \dots+x^{N-2})\}' \arrowvert_{x=2} = \\
\nonumber
= \frac{1}{4}\{x^2\frac{x^{N-1}-1}{x-1}\}' \arrowvert_{x=2} = \frac{1}{4}(\frac{x^{N+1}-x^2}{x-1})' \arrowvert_{x=2} = \frac{1}{4}2^N(N-1).
\end{eqnarray}
Now, we are interested in computing the \emph{average}, in step N, of values $\frac{n(N)}{N}$: we need to divide $\frac{1}{4}2^N(N-1)$ by $N$ and by the \emph{total} number of tree elements in step $N$, which is $1+2+\dots+2^{N-2} = \frac{2^{N-1}-1}{2-1} = 2^{N-1}-1$. That is, the average we look for is $\frac{1}{4}2^N(N-1)/N(2^{N-1}-1) = \frac{1}{4}(1-\frac{1}{N})\frac{2^N}{2^{N-1}-1}$; the limit value when $N$ grows is $\frac{1}{2}$. 
\subsection{Averaging $\sum_{j=1}^n \log(a_j+1)$ in step $N$}
We need now to repeat the process with magnitude $\sum_{j=1}^n \log(a_j+1)$ for all tree elements in step $N$; i.e. we need to sum all $\log (a_j+1)$ for each coefficient $a_j$ which appears in each tree element in step $N$, and then divide the sum by $N(2^{N-1}-1)$.
\newline
We start by considering, again, the restricted tree: see Fig. 1 (Diagram $D_1$). 
\begin{figure}
	\includegraphics[clip,height=2.5in,width=5.0in]{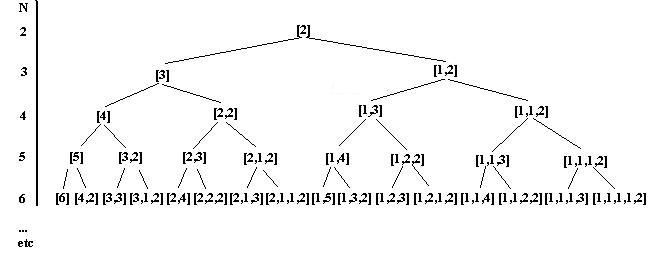} 
\caption{Diagram $D_1$}
\end{figure}
We notice that the sum of the coefficients $a_j$ inside a pair of brackets equals, exactly, the value of $N$ in which this tree element is located: the last one of the last row: $[1,1,1,1,2]$ fulfills $1+1+1+1+2=6=N$. We also recall the law shown in Fig. 2 (Diagram $D_2$).
\begin{figure}[tbh]
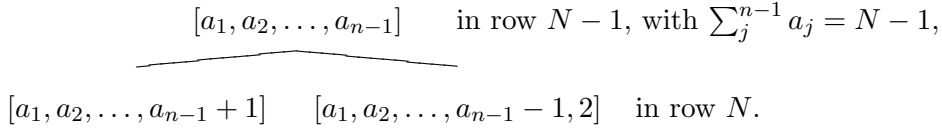
 
\xytree{
& & \xynode[-2,2]{$[a_1,a_2,\dots,a_{n-1}]$}& &&& & \xynode{in row $N-1$, with $\sum_j^{n-1} a_j = N-1$,} \\
\xynode{$[a_1,a_2,\dots,a_{n-1}+1]$} & & & & \xynode{$[a_1,a_2,\dots,a_{n-1}-1,2]$} & & & \xynode{in row $N$.}
}
\label{fig:D2}
\caption{Diagram $D_2$}
\end{figure} 
\newline
We start by counting the number of coefficients $a_k=1$ in step $N$. We observe that, when in a certain step a coefficient \textquotedblleft 1" appears, then it appears twice in the following step: e.g. 
\xytree{
& \xynode[-1,1]{$[1,3]$} \\
\xynode{$[1,4]$}  & & \xynode{$[1,2,2]$}
}
in steps $N=4$ and $N=5$ in the diagram $D_1$. Next, we observe that, when a coefficient $a_k=2$ appears as the last one in a tree element in a certain step, then it yields an $a_k=1$ in the following step: e.g. 
\xytree{
& \xynode[-1,1]{$[3,2]$} \\
\xynode{$[3,3]$}  & & \xynode{$[3,1,2]$}
}
in steps $N=5$ and $N=6$ in the same diagram. Therefore, the number of $a_k=1$ in step $N$ is: the double of the number of $a_k=1$ in step $N-1$, plus the number of last $2$'s in step $N-1$. But this last number, observing diagram $D_2$ inside $D_1$, is the number of elements in row $N-2$, i.e. $2^{N-4}$. In a notation that, we trust, is natural, we can write: $a_N(=1) = 2a_{N-1}(=1) + 2^{N-4}$. Starting from $N=3$, where $a_N(=1)=1$, we would have: $a_4(=1)=2\cdot 1+1$. In $N=2$ there is one $a_k=2$ which is the last coefficient (as well as the only one). So $a_4(=1)=2a_3(=1)+2^{4-4}=2+2^{4-4}$. Reiterating this law we have $a_5(=1)=2^2+2^{5-4}$ plus $2^{5-4}=2^2+2\cdot 2^{5-4}, a_6(=1)=2^3+3\cdot 2^{6-4} \dots $ and  $a_N(=1)=2^{N-3}+(N-3)2^{N-4}$ which satisfies the counting of $1$'s in diagram $D_1$ starting from $N=3$.
\newline
\indent
In order to obtain the total number of $1$'s in non-restricted $N$-step we have to add $2^{N-3}+(N-3)2^{N-4}$ from $N=3$ to $N=N$. We have $2^{N-3}+(N-3)2^{N-4}=2\cdot 2^{N-4}+(N-3)2^{N-4}=(2+N-3)2^{N-4}=(N-1)2^{N-4}= \frac{1}{4}(N-1)2^{N-2}=\frac{1}{4}(N-1)x^{N-2} \arrowvert_{x=2} = \frac{1}{4}(x^{N-1})'\arrowvert_{x=2}$, so the sum is $\frac{1}{4}(x^{3-1}+\dots+x^{N-1})'\arrowvert_{x=2} = \frac{1}{4}(x^2[1+\dots+x^{N-3}])'\arrowvert_{x=2} = \frac{1}{4}(x^2\frac{x^{N-2}-1}{x-1})'\arrowvert_{x=2}=(N-2)2^{N-3}$.
\newline
\indent
So $a_N(=1)=(N-2)2^{N-3}$; an equality valid from $N=2$ onwards. Here $a_N(=1)$ represents the number of $1$'s in the \emph{non}-restricted $N$-step.
\newline
\indent
From diagram $D_2$ inside $D_1$ we understand that it is a different problem to count the $a_N(=2)$. Again we start with restricted step $N$.
\begin{description}
    \item[A)] From diagram $D_2$ : 
\xytree{
& 	& \xynode[-2,2]{$[a_1 \dots a_{n-1}]$} \\
	\xynode{$[a_1 \dots a_{n-1}+1]$}& & & & \xynode{$[a_1 \dots a_{n-1}-1,2]$} \\	
}
\newline
we observe: \begin{enumerate}
    \item each tree term $[a_1 \dots a_{n-1}]$ introduces a number $2$ at the end of $[a_1 \dots a_{n-1}-1,2]$; and 
	\item from diagram $D_1$ we see that no number $a_{n-1}+1$ in $[a_1 \dots a_{n-1}+1]$ is $2$. So in step $N$ we have a number of \textquotedblleft ending 2's" equal to the total of tree elements in step $N-1$, i.e. $2^{N-3}$.
\end{enumerate} 
	\item[B)] Let us consider the case $[b_1\dots  b_k,2,2]$. Following the evolution that produces tree element $[1,1,2,2]$ in $D_1$ we observe :$[1,1,2] \rightarrow [1,1,3] \rightarrow [1,1,2,2]$, so, case $[b_1\dots  b_k,2,2]$ in step $N$ comes from the \textquotedblleft ending 2" two steps above\dots which in turn come from the total of tree elements another step above: $2^{N-5}$.
	\item[C)] The other $2$'s come from duplicating those in the step above:
\newline
\xytree{
& 	& \xynode[-2,2]{$[a_1,\dots,2,\dots,a_{n-1}]$} \\
	\xynode{$[a_1\dots,2,\dots,a_{n-1}+1]$}& & & & \xynode{$[a_1,\dots,2,\dots, a_{n-1}-1,2]$} \\	
}
\newline
With a procedure similar to the one for counting $1$'s this number is $2^{N-5}(N-3)$. 
\end{description}
\indent\indent
So the number $a_N(=2)$ of $2$'s in the $N$ step of the restricted tree is $2^{N-3}+2^{N-5}+(N-3)2^{N-5}=2^{N-5}(N-2)+2^2\cdot 2^{N-5}= 2^{N-5}(N+2)$. This formula works from $N=4$ onwards. The total number of $2$'s in $N=2$ and $N=3$ is $2$. So we need to sum $2^{N-5}(N+2)$ from $N=4$ to $N=N$ and add $2$ to this sum. We add $2^{N-5}(N+2)$ in exactly the same way in which we added $2^{N-4}(N-1)$, and we finally obtain that $a_N(=2)$ in the non-restricted $N$-step is $(N+1)2^{N-4}$.
\newline
\indent
To count $1$'s and $2$'s was a special problem, but ending $3$'s in step $N$ are produced by ending $2$'s in step $N-1$, in a natural way, and observing their evolution \textemdash following rules already laid out\textemdash\space in diagram $D_2$ inside $D_1$, we have $a_N(=3)=N2^{N-5}$ in non-restricted step $N$; similarly $a_N(=4)=(N-1)2^{N-6}$; $a_N(=5)=(N-2)2^{N-7}\dots$ and so on.
\newline
\indent
How do these quantities agree with a concrete \emph{finite} row in step $N$? 
\newline
\indent
Integer $1$ appears $(N-2)^{N-3}$ times; $2$ appears $(N+1)2^{N-4}$ times, whereas $3$ appears $N2^{N-5}$ times, 4 does $(N-1)2^{N-6}$ times \dots and $k$ appears $(N+3-k)2^{N-k-2}$ times if $k \geq 3.$ But we know that $k=N-1$ appears only twice in row $N$, and $k=N$ only once. How does $(N+3-k)2^{N-k-2}$ agree with these two quantities? For $k=N-1$ we obtain $(N+3-(N-1))2^{N-(N-1)-2}=4\cdot 2^{1-2}=\frac{4}{2}=2$ exactly, whereas, for $k=N$ the formula yields $(N+3-N)2^{N-N-2}=\frac{3}{4} \cong 1$: we are short by $0.25$, which does not affect our final result \textemdash it will be negligible when we divide the total sum by $N(2^{N-1}-1)$. With some care we have to sum now: $(N-2)2^{N-3}\log(1+1)+(N+1)2^{N-4}\log(2+1)+ N2^{N-5}\log(3+1)+(N-1)2^{N-6}\log(4+1)+\dots+(N+3-k)2^{N-(2+k)}\log(k+1)+ \dots 5\cdot 2^0\log([N-2]+1)+4\cdot 2^{-1}\log([N-1]+1)+ 3\cdot 2^{-2}\log(N+1)$ and divide this sum by $N$ and by $2^{N-1}$.
\newline
\indent
Let us add all except first and second terms above, rewriting the elements: $[N-(3-3)]2^{N-(3+2)}\log(3+1) + [N-(4-3)]2^{N-(4+2)}\log(4+1)+\dots+ [N-(k-3)]2^{N-(k+2)}\log(k+1)+\dots+[N-(N-3)]2^{N-(N+2)}\log(N+1)$. Now, we divide by $N$ and by $2^{N-1}$, and we obtain: $2^{-(3+1)}\log(3+1)+(1-\frac{1}{N})2^{-5}\log 5+\dots+(1-\frac{k-3}{N})2^{-(k+1)}\log(k+1)+\dots+(1-\frac{N-3}{N})2^{-(N+1)}\log(N+1) = 2^{-4}\log 4+ 2^{-5}\log 5+\dots+2^{-(k+1)}\log(k+1)+\dots+2^{-(N+1)}\log(N+1) - \frac{1}{N}\{2^{-5}\log 5 + \dots +(k-3)2^{-(k+1)}\log(k+1)+\dots+(N-3)2^{-(N+1)}\log(N+1)\}$. Let us consider the expression between brackets: $\sum_nA_n$, where $A_n=(n-3)2^{-(n+1)}\log(n+1)$. We have $\frac{A_{n+1}}{A_n} = \frac{n-2}{n-3}\frac{2^{-(n+2)}}{2^{-(n+1)}}\frac{\log(n+2)}{\log(n+1)} \rightarrow \frac{1}{2}$ if $n \rightarrow \infty$; so the expression in brackets is bounded as $N$ grows, so when we multiply it by $\frac{1}{N}$ and let $N\rightarrow \infty$ it vanishes. We are left with $2^{-4}\log 4+\dots+2^{-(N+1)}\log(N+1)$ as $N$ grows. Now, we had left aside two terms: $(N-2)2^{N-3}\log 2$ and $(N+1)2^{N-4}\log 3$, which we have to sum, and then divide by $N$ and by $2^{N-1}$. We obtain $(1-\frac{2}{N})2^{-2}\log 2+(1+\frac{1}{N})2^{-3}\log 3$ which, as $N$ grows, tends to $2^{-2}\log 2 +2^{-3}\log 3$. Finally, we are left with $2^{-2}\log 2+2^{-3}\log 3+\dots+2^{-(N+1)}\log (N+1)$ as $N$ grows, which is $\frac{1}{2}\{ \frac{\log 2}{2} +\frac{\log 3}{2^2} +\dots+\frac{\log(k+1)}{2^k}+\dots \}$. 
\newline
Adding all up we have the value in the denominator of $\frac{\log 2}{\log A}: 2\{ \frac{1}{2}\log c+\frac{1}{2}\{\frac{\log 2}{2}+\dots+\frac{\log(k+1)}{2^k}+\dots \}\}$ as $\log A$, which means $\log c+\frac{\log 2}{2^1}+\dots+\frac{\log(k+1)}{2^k}+\dots = \log A$; in the numerator we have $\log 2$ since the Farey tree is a left-right partition, like the ternary of Cantor.
\end{document}